\documentclass{aastex61}

\usepackage{amssymb}
\usepackage{multirow}
\usepackage{morefloats}
\usepackage{amsmath}
\usepackage{bigstrut}
\usepackage{booktabs}
\usepackage{float}
\usepackage{graphicx,epsfig,fancyhdr,epsf,txfonts,epstopdf}
\usepackage{latexsym,bbm}
\usepackage{lineno}
\usepackage{color}
\usepackage{ulem}
\usepackage{threeparttable}
\usepackage{longtable}

\received{10 July 2018}
\revised{30 August 2018}
\accepted{ZZZ}

\shortauthors{Li et al.}
\shorttitle{AGN variability from 3D relativistic jets}
\begin{document}

\title{Radio-loud AGN variability from three-dimensional propagating relativistic jets}

\correspondingauthor{Paul J. Wiita}
\email{wiitap@tcnj.edu}

\author{Yutong Li}\thanks{E-mail: qiuwutongyu622@163.com, liy@tcnj.edu}
\affil{Shandong Provincial Key Laboratory of Optical Astronomy and Solar-Terrestrial Environment, Institute of Space Sciences, \\
Shandong University, Weihai, 264209, China\\}
\affiliation{Department of Physics, The College of New Jersey, 2000 Pennington Rd., Ewing, NJ 08628-0718, USA}

\author{Paul J. Wiita}\thanks{E-mail: wiitap@tcnj.edu}
\affiliation{Department of Physics, The College of New Jersey, 2000 Pennington Rd., Ewing, NJ 08628-0718, USA}

\author{Terance Schuh}
\affiliation{Department of Physics, The College of New Jersey, 2000 Pennington Rd., Ewing, NJ 08628-0718, USA}

\author{Geena Elghossain}
\affiliation{Department of Physics, The College of New Jersey, 2000 Pennington Rd., Ewing, NJ 08628-0718, USA}

\author{Shaoming Hu}
\affiliation{Shandong Provincial Key Laboratory of Optical Astronomy and Solar-Terrestrial Environment, Institute of Space Sciences, \\
Shandong University, Weihai, 264209, China\\}


\begin{abstract}

\noindent
The  enormous sizes and variability of emission of radio-loud AGNs arise from the relativistic flows of plasma along two oppositely directed jets. We use the Athena hydrodynamics code to simulate  an extensive suite of 54 propagating three-dimensional relativistic jets with  wide ranges of input jet velocities and jet-to-ambient matter density ratios.  We determine which parameter sets yield unstable jets that produce jet dominated FR I type radio galaxy morphologies  and which tend to produce stable jets with hot-spots and FR II morphologies.   Nearly all our simulations involve jets with internal pressures matched to those of the ambient medium but we also consider over-pressured jets and discuss differences from the standard ones. We also show that the results are not strongly dependent on the adiabatic index of the fluid.  We  focus on  simulations that remain  stable for extended distances (60 -- 240) times the initial jet radius. Scaled to the much smaller sizes probed by VLBI observations, the fluctuations in such simulated  flows yield variability in observed emissivity on timescales from months.  Adopting results for the densities, pressures and velocities from these  simulations we  estimate normalized rest frame synchrotron emissivities from individual cells in the jets. The observed emission from each cell is strongly dependent upon its variable Doppler boosting factor.  We  sum the  fluxes from thousands of zones around the primary reconfinement shock. The light curves and power-spectra, with red-noise slopes between $-2.1$ and $-2.5$, so produced are similar to those observed from blazars.
\end{abstract}

\keywords{galaxies: active -- BL Lacertae objects: general -- galaxies: jets -- quasars: general}

\section{Introduction}

Active galactic nuclei (AGNs) are powered by accretion onto supermassive black holes in the centers of their host galaxies.
Around 10 percent of AGNs are classified as radio-loud \citep[e.g.][]{Jiang07}, and are characterized by relativistic plasma jets, the extended linear structures that transport energy and particles from the compact central region out to kpc or even Mpc-scale extended regions \citep{Begel84,Peter97}.
Variability in observed emission can be considered a defining characteristic of AGNs, and for these radio-loud AGN the majority of this variable emission is understood to arise from synchrotron emission from relativistic flows of plasma along two oppositely directed jets \citep{Urry95}.  The combination of multi-band radio observations and theoretical modeling has provided extremely strong evidence for the presence of both moving and standing shocks in these jets \citep[e.g.][]{Aller85,Mars85,Hughes91,Lister01,Lister09,Mars08,Mars10}.
It is now generally accepted that the largest flares arise from the production and relativistic propagation of new components seen as radio knots \citep[e.g.][]{Hughes91} but smaller observed variations can arise from changes in the overall directions, and hence Doppler factors, of the jets \citep[e.g.][]{Camenzind92,Gopal92}, turbulence within the jets \citep[e.g.][]{Mars08,Mars14,Calafut15}, or differential velocities between small regions within the jets \citep[e.g.][]{Giannios09,Poll16}.
Changes in the overall direction of the inner portions of the  jet
have clearly been demonstrated using very long baseline interferometry (VLBI) \citep[e.g.][]{Bire86,Ros00,Piner03,Caproni04,Lister13}.  Blazars, the set of AGNs that is the union of BL Lacertae objects and Flat Spectrum Radio Quasars (or at least the high-polarization subset of FSRQs) are highly variable throughout the electromagnetic spectrum; their enhanced brightness and strong variability are understood to arise through Doppler boosting when one of the jets is pointing close to our line-of-sight \citep[e.g.][]{Urry95}.

Extended extragalactic radio sources have long been classified into two categories \citep{Fan74} based upon their radio morphology.  Fanaroff-Riley I (FR I) sources haves jet-dominated emission, with the majority of their fluxes arising from the inner halves of their total extent and  are weaker.
The FR II, or classical double sources, have emission dominated by lobes containing terminal hot-spots; frequently only one of the jets feeding those lobes are detected.  They can be distinguished based on power: objects below $\sim 10^{25} h_{70}^{2} {\rm W~Hz}^{-1}{\rm str}^{-1}$ at 178 MHz were typically found to have FR I morphologies (where $h_{70}$ is the Hubble constant in units of 70 km s$^{-1}$ Mpc$^{-1}$). An examination of the radio luminosity
at 1.4 GHz against the optical absolute magnitude of the host galaxy led to the finding \citep{Owen94,Ledlow96} that the radio flux boundary line between FR I and FR II sources
correlates as $L_{radio}\propto L_{optical}^{1.7}$, which means that more radio power is required to form a FR II radio source in a more luminous galaxy.
Furthermore, some hybrid-morphology radio sources (HYMORS) have been discovered that show FR I structure on one side of the radio source and FR II morphology on the other \citep{Gopal00}. These sources are important in understanding the basic origin of the FR I and FR II dichotomy, where the different morphologies may be induced by intrinsically different jet properties, interactions with different environments on either side of the source \citep[e.g.][]{Bick94,Gopal01,Mass16}, or long-term temporal variations combined with the time-lag in the observer's frame between evolving approaching and receding lobes \citep[e.g.][]{Gopal00,Gopal04,Mass03,Wold07,Kawa09,Kapinska17}.

Hydrodynamical simulations of propagating jets are of critical importance to the understanding of radio galaxies and quasars and now have a history spanning four decades  \citep{Ray77,Wiita78,Norman82}.
These simulations give fundamental support to the idea of the twin-jet models for radio galaxies dating to \cite{Scheuer74} and \cite{Bland74}.
Two basic parameters, the internal beam Mach number, $M_{b}$, and the jet to ambient density contrast, $\eta$, were found to govern the morphology and propagation properties of these jets \citep{Norman82}.  Good numerical and analytical understandings of 2D and 3D non-relativistic hydrodynamical flows were achieved by the mid-1990s \citep[e.g.][]{Hardee88,Hardee90,Burns91,Hooda94,Bass95,Bick95,Bodo95}.  Since then, the complexity of the simulations has increased  in parallel with growing computational power and algorithm development, leading to better understanding of the jet phenomenon, focussing on the study of the morphology, dynamics and non-linear stability of jets at kpc and larger scales \citep[e.g.][]{Norman96,Fer98,Car02a,Car02b,Krause07,Dono16,Mass16}.

Early simulations of special relativistic jets were carried out by a few groups.  Some focused on the propagation on nuclear regions corresponding to the parsec-scale structures revealed by VLBI \citep[e.g][]{Marti95,Marti97,van96,Miod97}, while others focused on propagation out to larger scales and production of the structures seen on the multi-kiloparsec scale of powerful radio galaxies \citep[e.g.][]{Rosen99,Hardee01,Marti03,Leis05,Mass16}.

Of course, extragalactic jets emit by the synchrotron mechanism (at least from radio through ultraviolet bands) and equipartition arguments have long indicated that the energy density in magnetic fields should be substantial.  Nonetheless, there have been a more modest number of simulations that include them dynamically in magnetohydrodynamical (MHD) codes, particularly in 3D \citep[e.g.][]{Clarke86,Nishi97,Leis05,ONeill05,Keppens08,Gaibler09,Mignone10,Huarte11,Hard14,Marti15,Marti16}.  The relative paucity of work in this area can be attributed to both the difficulty of maintaining conservation of magnetic flux and energy in MHD codes (especially special relativistic ones) and to the early indications that simulations of jets with dynamically important magnetic fields did not resemble observed radio source morphologies \citep{Clarke86}.  We shall not discuss MHD jet simulations further in this paper, but note that the Athena code we use is designed to handle them and that we are in the process of producing a suite of such simulations to be compared to the relativistic HD ones presented here.

The Athena code, developed by J.\ Stone and colleagues \citep{Gard05,Stone08,Beck11} is an excellent choice for a wide range of astrophysical computational fluid dynamics problems.
It is a highly efficient, grid-based, code for astrophysical MHD that was developed primarily for studies of the interstellar medium, star formation, and accretion flows. Athena has the capability to include special relativistic hydrodynamics (RHD) and MHD, static (fixed) mesh refinement and parallelization using domain decomposition and MPI. The discretization is based on cell-centered volume averages for mass, momentum, and energy, and face-centered area averages for the magnetic field. In order to solve a series of discretized partial differential equations expressing conservation laws, the rest density, pressure, velocity, internal energy, and magnetic field are calculated though in these RHD simulations the
magnetic field was set to zero.

 Here we simulate propagating jets with a very extensive suite of both medium-power and high-power jets that are run for substantial problem times using larger computational resources.  These allow us to better distinguish between parameters likely to produce jets that remain stable for extended distances (FR II type) from those that do not (FR I type).  The other key aspect of this work is that we examine the light curves and power spectra from approximated jet emission taking into account the different velocities of the cells within the jet.  Although the literature on  simulations of jets for extragalactic radio sources is now extensive, such computations of light curves \citep[e.g.][]{Mars14} and the resulting power-spectra produced by the jet motions \citep{Calafut15,Poll16} are exceedingly rare.   Here we consider emission variations that are produced by changing Doppler factors, densities and pressures from a fixed region in a propagating jet.

However, it has been noted \citep{Poll16}, that faster sub-grid variability is likely to involve relativistic turbulence  \citep[e.g.][]{Zrake13} which is almost certainly present with these jets, at least in some portions \citep{Mars08,Mars10}.  \cite{Mars14} has created a sophisticated ``turbulent extreme multi-zone'' model that posits the presence of turbulence and computes beautiful multi-band light curves for flows through a standing-shock within a jet of fixed bulk velocity.  However, neither this TEMZ model, nor the simpler turbulent model of \cite{Calafut15},
take into account the variability arising from the bulk jet motions.  Our group has previously performed relativistic jet simulations, albeit in only two dimensions, using the Athena code in \cite{Poll16}, where wiggles in slab jets were computed in conjunction with a simplified turbulence model.  The jet velocity and density changes produced slower variations while the turbulence produced faster ones that connected with them; these models had the advantage of producing variations over five orders of magnitude in time, and with power spectral slopes similar to those usually observed in AGN.  There the changes in observed   synchrotron emission were estimated by summing the fluxes from a vertical strip of zones behind the reconfinement shock and  power spectral densities (PSDs) were calculated from the light curves for both turbulent and bulk velocity origins for variability; these yielded PSDs  with slopes in the ranges $-1.8$ to $-2.3$ and  $-2.1$ to $-2.9$, respectively \citep{Poll16}.   Here we extend the bulk calculations to 3D and include orders of magnitude more zones in the light curve estimations.

In Section 2 we provide a description of our RHD simulations.   Our basic results are given in Section 3 for an exceptionally large number of jet simulations with a substantial range of jet velocities and densities; there we consider typical jet evolutions and compare higher and lower power inputs that respectively give rise to stable and unstable jets.  In Section 4 we describe how we compute reasonable estimates for the light curves, focusing on the portion of the jet near the first reconfinement shock, where both intrinsic emissivity and Doppler boosting are high.  In Section 5 we consider differences between pressure-matched and over-pressured jets, the effects of different adiabatic indices and give a brief comparison of 2D and 3D jets computed with Athena.  In Section 6 we present and discuss light curves and power spectral densities (PSDs) for representative simulations.  Our conclusions are given in Section 7.

\section{Simulations of 3D RHD jets}

 We use the Athena code \citep{Gard05,Stone08,Beck11} for special relativistic hydrodynamics to produce 3D simulations of  jets propagating through initially uniform external, or ambient, media, with wide ranges of powers.   A relativistic hydrodynamic jet problem in three-dimensions is available in the problem files bundled with Athena\footnote{\url{https://trac.princeton.edu/Athena}}.  Our jets are computed as launched along the $x$-axis in cartesian coordinates, and the circular jet inlet is taken to have a radius $R_j = 1$ unit (so that all lengths can be scaled to that unit) in the $x = 0$ plane.   To model a hydrodynamic jet, the initial physical parameters of inlet jet velocity, $v_j$ (assumed constant across the inlet cross-section for our initially cylindrical jets),  proper (rest) ambient and jet densities ($\rho_a$ and $\rho_j$, respectively), ambient and jet pressures ($P_a$ and $P_j$) and adiabatic index, $\Gamma$, must be specified.   In our production runs we normally use $\Gamma = 5/3$ but we have performed a few runs with $\Gamma = 4/3$ and will discuss the differences so produced.  However, as has been shown in earlier work, the dominant variables are $v_j$ and $\eta = \rho_j / \rho_a$.  The great majority of our simulations are performed for pressure-matched jets ($P_a = P_j$) but we also consider a few over-pressured jets, with  $P_j = 10 P_a$.  Boundary conditions are set to outflow everywhere except at the inflowing jet inlet.  We used a Courant number of 0.4 in our runs except when smaller values were needed for the code to function successfully.  The constant density ambient media we model here are good approximations in both the nuclear region of a galaxy (scales $10 - 500$ pc) and on extended motions in intracluster media (ICM, on scales $40 - 400$ kpc).  However, studies of such jets in media with decreasing densities, as would be appropriate for propagation through galactic halos (on 1 to 10 kpc scales) are also important \citep[e.g.][]{Gopal91,Hooda98,Car02b,Jeya05,Mass16} and we also neglect the likely important impact of large-scale motions in the ICM on propagating jets \citep[e.g.][]{Loken95,Bliton98}.
The Athena code permits the user to optimize the simulation by allowing a choice of the order of reconstruction (always taken as second-order piecewise parabolic in our production runs), as well as providing options such as
 static mesh refinement which can change the resolution of a grid by a factor of two  per refinement level. This can allow for more focus on parts of the grid that contain interesting phenomena, such as the jet or shocks, although we did not employ mesh refinement.

 With substantial experimentation involving different code parameters, our best overall results for faster jets came from simulations of higher resolution 3D RHD jets with $1200 \times 500 \times 500$ zones  with the  length along the axis extending to 120  $R_{j}$, and on $ 600 \times 500 \times 500$ for grids extending only out to 60 $R_j$ (or 10 zones in each length unit, hence 20 zones across the jet diameter).    However, many preliminary runs  and some production runs at those lengths were performed at a medium resolution (5 zones in each unit, or 10 across the jet diameter) including one with the jet propagating an extremely long distance (for 3D simulations, anyway) of  240 $R_{j}$.  For $R_{j} = 500$ pc, as is reasonable for jets propagating on extragalactic scales, one time unit equals 1630 yr.  However, if we consider these simulated jets to be propagating on VLBI scales, a reasonable value for $R_j$  might be 1 pc and a time unit in the source rest frame then would be 3.26 yr.  These domain extensions, $L_{x}\times L_{y} \times L_{z}$, and numbers of cells in each dimension,  $N_{x}\times N_{y}\times N_{z}$, for our models are given in Table \ref{tab:parameter} along with the corresponding values of jet velocities $\beta_{j}$, Lorentz factor $\gamma_{j}$,  and jet-to-ambient density ratios $\eta$.  Also given in Table \ref{tab:parameter} are the equivalent Mach number $M_c$, jet luminosity $L_j$, and the imposed limit time of each run $T_1$. Results given in this table are the times that the jets start to become unstable $T_2$, the distances that jets retain stability $S_j$, whether or not the jets reach the ends of the grids (``End''),  and the expected FR type.  We also computed three overpressured simulations with ambient (dimensionless) pressure $P_a = 10^{-3}$ while the jet pressure $P_j = 10^{-2}$  ($P_j / P_a =10$); the default pressure values are  $P_a = P_j = 10^{-2}$).  The three simulations that were also performed with  overpressured jets are marked with asterisks in Table \ref{tab:parameter} and those also performed with a different adiabatic index are noted with daggers.

\begin{figure*}
    \centering
    \includegraphics [trim=0.2cm 3cm 0.2cm 5cm,width=0.85\textwidth,clip]{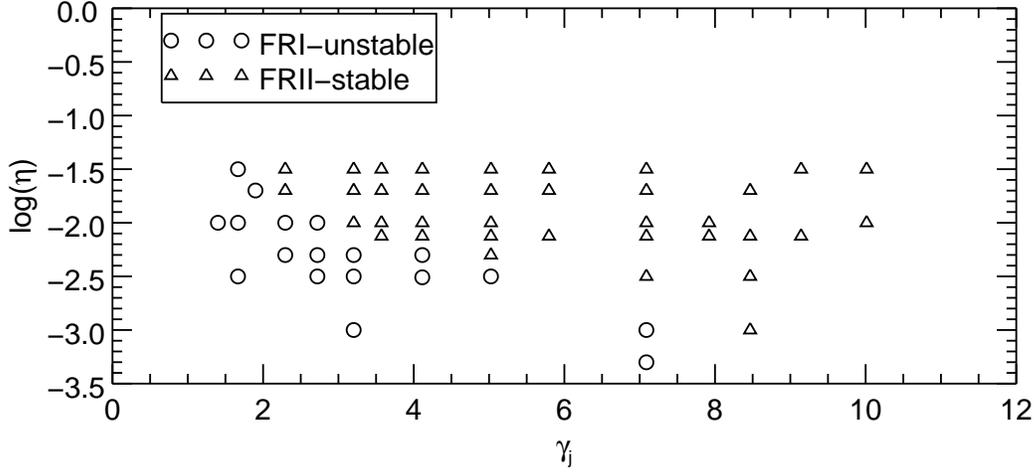}
    \caption{Summary of the stability (and likely morphology) of our 54 simulations.  Circles represent FR I runs which have unstable jets; triangles are FR II runs with stable jets.}
    \label{fig:datatable}
 \end{figure*}

A total of 54 RHD simulations have been performed with the Athena code with different combinations of jet velocities ($v_{j}$) and jet-to-ambient matter density ratios ($\eta$). The simulations contain a range of $\eta$ from 0.0005 to 0.0316 and a range of initial $v_{j}$ from 0.7c ($\gamma_j = 1.4$) to 0.995c ($\gamma_j = 10$), where, as usual, the Lorentz factor is $\gamma_j = [1-\beta_j^2]^{-1}$ with $\beta_j = v_j/c$.  A summary of the results of these simulations are shown in Figure \ref{fig:datatable}. The circles in Figure \ref{fig:datatable} represent runs with jets that eventually go unstable before the end of the grid at 60 or 120 $R_j$ is reached, and can be considered to correspond to FR I radio sources if scaled to extragalactic dimensions. Triangles show parameters of runs with jets that remain stable enough to retain terminal shocks throughout the simulation (to even 240 $R_j$) and thus are plausibly representative of FR II sources. The boundary between stable and unstable jets corresponds nicely to their  nominal powers, as will be discussed further in Section 5.1.

One can compare Newtonian jets and special relativistic jets through defining $\eta_{c}$ and $M_{c}$ as the equivalent classical density ratios and Mach numbers, respectively \citep{Rosen99}. In the formulation of \cite{Car02a},
  \begin{equation}
   \eta_{c} = \gamma_{j}^{2}\eta (1+\delta_{2}),
  \label{equ:etac}
  \end{equation}
  \begin{equation}
   M_{c}=\sqrt{3} \gamma_{j}\beta_{j}(1+\delta_{2}^{-1})^{1/2}.
  \label{equ:Mc}
  \end{equation}
 Here $\delta_{2}$ is a factor related to the energy density of relativistic electrons, and is defined as
  \begin{equation}
   \delta_{2}=\frac{4}{3} (1+k) \overline{\gamma_{e}} \frac{m_{e}}{m_{p}}q ,
  \label{equ:deta2}
  \end{equation}
where $k\simeq 0$ for a pair plasma jet (Case 1, \cite{Car02a}), while $k = 100$, neglecting the thermal component, is typically taken for a proton-electron jet (Case 2); here $\overline{\gamma_{e}}$ is the average Lorentz factor of the relativistic electrons, and $q$, the ratio of the number of
electrons to total particles within the jet, is taken to be 0.5. For a limiting fully relativistic adiabatic index $\Gamma = 4/3$, $\delta_{2}$ is 0.106 (Case 1 in \cite{Car02a}).  For our calculations, we have nearly always employed a value of $\Gamma = 5/3$ appropriate for an ionized monatomic gas; then $\delta_{2} \simeq 2.75$, in approximate accordance with $k\simeq 100$.
We include $M_c$ in Table \ref{tab:parameter}.

The following expression gives the energy flux of a jet where relativistic particles are important  \citep{Bick94,Bick95}
\begin{equation}
L_j = 4\pi c ~P_j ~R_j^2 ~[1 + (\gamma_j -1)\zeta /\gamma_j]~ \gamma_j^2 ~\beta_j,
 \label{equ:jetpower}
\end{equation}
where $P_j$ is the jet pressure and $\zeta = \rho_j c^2 / (\epsilon + P_j)$ is the ratio of cold matter energy density to enthalpy and $\epsilon$ is the internal energy density; when the internal energy and pressure are dominated by relativistic particles then $\zeta = \rho_j c^2 / 4P_j$ \citep{Bick95}. We use this expression, taking a fully-ionized gas mixture of hydrogen and helium so the mean molecular weight,  $\mu=0.60$ and then $\rho_j = \eta \mu m_p n_{a}$, where $n_a$ is the number density of the ambient gas.  In the limit where the cold material dominates over the relativistic particles, Equation \ref{equ:jetpower} reduces to
\begin{equation}
L_j \simeq \pi c^3 ~R_j^2~ \eta ~\rho_{a} (\gamma_j -1)~ \gamma_j ~\beta_j.
\label{equ:approxpower}
\end{equation}

The simulations are carried out using dimensionless variables. However, it is sensible to present the results with scalings to physical units in order to facilitate comparison with astrophysical radio sources. Thus, if we consider large-scale jets on scales $\sim 100$ kpc propagating through a uniform intracluster medium (ICM)  we can take $ R_{j} = 0.5$ kpc and can take ICM number densities, temperatures and pressures to have typical values around $n_{ICM} = 1 \times 10^{-3} {\rm cm}^{-3}$, $T_{ICM} = 5.0 \times 10^7$ K and thus $P_{ICM} =   6.9 \times 10^{-12}$ dyne \citep{Sara88}. The values of $L_j$ for these parameters are given in Table \ref{tab:parameter},  where a range from $5.82 \times 10^{44}$ erg s$^{-1}$ to $5.73 \times 10^{47}$ erg s$^{-1}$ is spanned.  In Section 4 we consider variability for which  jets propagating through the nuclear core of the galaxy (which is also nearly constant in density inside the core radius) are relevant instead; then,  a sensible scale is set by taking $R_{nuc} = 1$~pc and $n_{nuc} \approx 1 {\rm cm}^{-3}$ for the ambient gas.  The equivalent jet powers for the nuclear region are essentially reduced by a factor of $(R_{nuc}/R_{ICM})^2 (n_{nuc}/n_{ICM}) \approx 4 \times 10^{-3}$ when compared to the large-scale extragalactic ones for the same parameters.

\newpage

\begin{longtable}{lcccccccccccc}
  \caption{Simulation parameters}
  \label{tab:parameter}\\
    \hline\hline
     $\beta_{j}$& $\gamma_{j}$ & $\eta$ &$T_1^{\ 1}$& $L_{x}\times L_{y} \times L_{z}^{\ 2}$ & $N_{x}\times N_{y}\times N_{z}^{ \ 3}$& $CFL ^{\ 4}$&$M_{c}$&$L_{j}^{\ 5}$& $T_2 ^{\ 6}$& $S_{j} ^{\ 7}$& $End ^{\ 8}$ & Type \\\hline
       0.7  & 1.40    &  0.01 &1000 & $60\times 50 \times 50$ & $600\times 500 \times 500$ & 0.4 & 1.98 & 8.05(44)  & 700  & 24.0  & yes   & FR I     \\
       0.8  & 1.67    &0.00316&1100 & $60\times 50 \times 50$ & $600\times 500 \times 500$ & 0.4 & 2.70 & 5.82(44)  & 550  & 18.5  & no   & FR I     \\
       0.8  & 1.67    &  0.01 &1300 & $60\times 50 \times 50$ & $600\times 500 \times 500$ & 0.4 & 2.70 & 1.82(45)  & 325  & 25.8  & yes & FR I     \\
       0.8  & 1.67    & 0.0316&800  & $60\times 50 \times 50$ & $600\times 500 \times 500$ & 0.4 & 2.70 & 5.71(45)  & 275  & 36.8  & yes & FR I     \\
       $0.85^{\dagger}$& $1.90^{\dagger}$& $0.02^{\dagger}$&400& $60\times 50 \times 50$ & $600\times 500 \times 500$ & 0.4 & 3.27& 5.90(45)& 240&32.7&yes&FR I   \\
       0.9  & 2.29    & 0.005 &1000 &$120\times 50 \times 50$ & $600\times 250 \times 250$ & 0.4 & 4.17 & 2.75(45)  & 350  & 58.7  & no   & FR I     \\
       0.9  & 2.29    & 0.01  &1000 &$120\times 50 \times 50$ & $600\times 250 \times 250$ & 0.4 & 4.17 & 5.45(45)  & 425  & 64.2  & yes & FR I     \\
       0.9  & 2.29    & 0.02  &600  &$120\times 50 \times 50$ & $600\times 250 \times 250$ & 0.4 & 4.17 & 1.09(46)  & 400  & 111.5 & yes & FR I    \\
       0.9  & 2.29    & 0.0316&300  & $60\times 50 \times 50$ & $300\times 250 \times 250$ & 0.4 & 4.17 & 1.72(46)  & --   & $>60$ & yes & FR II    \\
       0.93 & 2.72    &0.00316&500  & $60\times 50 \times 50$ & $600\times 500 \times 500$ & 0.3 & 5.12 & 2.83(45)  & 100  & 16.6  & no   & FR I     \\
       0.93 & 2.72    & 0.005 &500  & $60\times 50 \times 50$ & $600\times 500 \times 500$ & 0.4 & 5.12 & 4.45(45)  & 225  & 31.6  & no   & FR I     \\
       0.93 & 2.72    & 0.01  &500  & $60\times 50 \times 50$ & $600\times 500 \times 500$ & 0.4 & 5.12 & 8.88(45)  & 240  & 51.2  & no   & FR I     \\
       0.95 & 3.20    &0.001  &1100 &$120\times 50 \times 50$ & $600\times 250 \times 250$ & 0.3 & 6.15 & 1.35(45)  & 300  & 31.1  & no   & FR I     \\
       0.95 & 3.20    &0.00316&600  & $60\times 50 \times 50$ & $600\times 500 \times 500$ & 0.4 & 6.15 & 4.35(45)  & 170  & 31.5  & no   & FR I     \\
       0.95 & 3.20    &0.005  &400  &$120\times 50 \times 50$ & $600\times 250 \times 250$ & 0.4 & 6.15 & 6.85(45)  & 350  & 57.7  & no   & FR I     \\
       0.95 & 3.20    &0.01   &500  &$120\times 50 \times 50$ & $600\times 250 \times 250$ & 0.4 & 6.15 & 1.37(46)  & --   & $>120$& yes & FR II    \\
       0.95 & 3.20    &0.02   &500  &$120\times 50 \times 50$ & $1200\times 500 \times 500$& 0.4 & 6.15 & 2.73(46)  & --   & $>120$& yes & FR II    \\
       0.95 & 3.20    &0.0316 &500  &$120\times 50 \times 50$ & $600\times 250 \times 250$ & 0.4 & 6.15 & 4.30(46)  & --   & $>120$& yes & FR II    \\
       0.96 & 3.57    &0.0075 &400  & $60\times 50 \times 50$ & $600\times 500 \times 500$ & 0.4 & 6.93 & 1.35(46)  & --   & $>60$ & yes & FR II    \\
       0.96 & 3.57    &0.01   &400  & $60\times 50 \times 50$ & $600\times 500 \times 500$ & 0.4 & 6.93 & 1.80(46)  & --   & $>60$ & yes & FR II    \\
       $0.96^{* \dagger}$& $3.57^{* \dagger}$&$0.02^{* \dagger}$&200 &$60\times 50 \times 50$ &$600\times 500\times 500$& 0.4 & 6.93&3.58(46) & --& $>60$ & yes&FR II \\
       0.96 & 3.57    &0.0316 &300  &$120\times 50 \times 50$ & $1200\times 500 \times 500$& 0.4 & 6.93 & 6.15(46)  & --   & $>120$& yes & FR II    \\
       0.97 & 4.11    &0.00316&800  &$120\times 50 \times 50$ & $600\times 250 \times 250$ & 0.2 & 8.06 & 7.90(45)  & 350  & 61.5  & yes & FR I     \\
       0.97 & 4.11    &0.005  &550  &$120\times 50 \times 50$ & $600\times 250 \times 250$ & 0.4 & 8.06 & 1.27(46)  & 380  & 80.8  & yes & FR I     \\
       0.97 & 4.11    &0.0075 &500  &$120\times 50 \times 50$ & $600\times 250 \times 250$ & 0.4 & 8.06 & 1.90(46)  & --   & $>120$& yes & FR II    \\
       0.97 & 4.11    &0.01   &500  &$120\times 50 \times 50$ & $600\times 250 \times 250$ & 0.4 & 8.06 & 2.53(46)  & --   & $>120$& yes & FR II    \\
       0.97 & 4.11    &0.02   &1000 &$120\times 50 \times 50$ & $600\times 250 \times 250$ & 0.4 & 8.06 & 5.05(46)  & --  & $>120$ & yes & FR II    \\
       0.97 & 4.11    &0.0316 &500  &$120\times 50 \times 50$ & $600\times 250 \times 250$ & 0.4 & 8.06 & 7.98(46)  & --  & $>120$ & yes & FR II    \\
       0.98 & 5.03    &0.00316&600  &$120\times 50 \times 50$ & $600\times 250 \times 250$ & 0.3 & 9.97 & 1.27(46)  & 400  & 80.7 & yes & FR I    \\
       0.98 & 5.03    &0.005  &600  &$120\times 50 \times 50$ & $600\times 250 \times 250$ & 0.2 & 9.97 & 2.01(46)  & --  & $>120$ & yes & FR II    \\
       0.98 & 5.03    &0.0075 &800  &$120\times 50 \times 50$ & $600\times 250 \times 250$ & 0.4 & 9.97 & 3.03(46)  & --  & $>120$ & yes & FR II    \\
       $0.98 ^{\dagger}$& $5.03^{\dagger}$&$0.01^{\dagger}$&400& $60\times 50 \times 50$ & $600\times 500 \times 500$ & 0.4 &9.97 & 4.03(46)& --& $>60$& yes&FR II   \\
       $0.98^{*}$& $5.03^{*}$&$0.02^{*}$&200 &$60\times 50 \times 50$ & $600\times 500 \times 500$& 0.4 & 9.97  &8.05(46) & -- & $>60$ & yes &FR II \\
       0.98 & 5.03    &0.0316 &400  &$120\times 50 \times 50$ & $600\times 250 \times 250$ & 0.4 & 9.97 & 1.27(47)  & --  & $>120$ & yes & FR II    \\
       0.985& 5.80    &0.0075 &300  &$120\times 50 \times 50$ & $600\times 250 \times 250$ & 0.4 &11.56 & 4.18(46)  & --  & $>120$ & no & FR II    \\
       0.985& 5.80    &0.02   &500  &$240\times 100 \times 100$ & $1200\times 500 \times 500$ & 0.4 &11.56 & 1.11(47)  & --  & $>240$ & yes & FR II    \\
       0.985& 5.80    &0.0316 &500  &$120\times 50 \times 50$ & $600\times 250 \times 250$ & 0.3 &11.56 & 1.75(47)  & --  & $>120$ & yes & FR II \\
       0.99 & 7.09    &0.0005 &600  &$120\times 50 \times 50$ & $600\times 250 \times 250$ & 0.1 &14.20 & 4.32(45) & 270  & 52.5  & no  & FR I    \\
       0.99 & 7.09    &0.001  &600  &$120\times 50 \times 50$ & $600\times 250 \times 250$ & 0.1 &14.20 & 8.63(45) & 280  & 57.9  & no  & FR I    \\
       0.99 & 7.09    &0.00316&500  &$120\times 50 \times 50$ & $600\times 250 \times 250$ & 0.3 &14.20 & 2.73(46)  & --  & $>120$  & yes  & FR II    \\
       $0.99^{*}$& $7.09^{*}$&$0.0075^{*}$&200&$60\times 50 \times 50$ & $600\times 500 \times 500$ & 0.4 &14.20&6.53(46)& --  & $>60$& yes & FR II \\
       0.99 & 7.09    &0.01   &150  & $60\times 50 \times 50$ & $300\times 250 \times 250$ & 0.4 &14.20 & 8.70(46)  & --  & $>60$  & yes  & FR II    \\
       0.99 & 7.09    &0.02   &400  &$120\times 50 \times 50$ & $600\times 250 \times 250$ & 0.3 &14.20 & 1.73(47)  & --  & $>120$  & yes  & FR II    \\
       0.99 & 7.09    &0.0316 &400  &$120\times 50 \times 50$ & $600\times 250 \times 250$ & 0.3 &14.20 & 2.73(47)  & --  & $>120$  & yes  & FR II    \\
       0.992& 7.92    &0.0075 &300  &$120\times 50 \times 50$ & $600\times 250 \times 250$ & 0.4 &15.89 & 8.30(46)  & --  & $>120$ & yes  & FR II    \\
       0.992& 7.92    &0.01   &300  &$120\times 50 \times 50$ & $600\times 250 \times 250$ & 0.4 &15.89 & 1.11(47)  & --  & $>120$ & yes  & FR II    \\
       0.993& 8.47    &0.001  &500  &$120\times 50 \times 50$ & $600\times 250 \times 250$ & 0.2 &17.01 & 1.27(46)  & --  & $>120$ & yes  & FR II    \\
       0.993& 8.47    &0.00316&400  &$120\times 50 \times 50$ & $600\times 250 \times 250$ & 0.3 &17.01 & 4.23(46)  & --  & $>120$ & yes  & FR II    \\
       0.993& 8.47    &0.0075 &300  &$120\times 50 \times 50$ & $600\times 250 \times 250$ & 0.4 &17.01 & 9.58(46)  & --  & $>120$ & yes  & FR II    \\
       0.993& 8.47    &0.02   &300  &$120\times 50 \times 50$ & $600\times 250 \times 250$ & 0.3 &17.01 & 2.55(47)  & --  & $>120$ & yes  & FR II    \\
       0.994& 9.14    &0.0075 &300  &$120\times 50 \times 50$ & $600\times 250 \times 250$ & 0.3 &18.38 & 1.12(47)  & --  & $>120$ & yes  & FR II    \\
       0.994& 9.14    &0.0316 &300  &$120\times 50 \times 50$ & $600\times 250 \times 250$ & 0.2 &18.38 & 4.73(47)  & --  & $>120$ & yes  & FR II    \\
       0.995& 10.01   &0.01   &300  &$120\times 50 \times 50$ & $600\times 250 \times 250$ & 0.3 &20.15 & 1.81(47)  & --  & $>120$ & yes  & FR II    \\
       0.995& 10.01   &0.0316 &300  &$120\times 50 \times 50$ & $600\times 250 \times 250$ & 0.1 &20.15 & 5.73(47)  & --  & $>120$ & yes  & FR II    \\
    \hline
    \end{longtable}

\begin{tablenotes}
\item[1] 1. $T_1$ is the prescribed time limit of the simulation.
\item[2] 2. $L_{x}\times L_{y} \times L_{z}$ gives the extent of the domain in units of $R_{j}$.
\item[3] 3. $N_{x}\times N_{y}\times N_{z}$ means numbers of grid points in the three directions.
\item[4] 4. The Courant, Friedrichs, \& Lewy (CFL) Number
\item[5] 5. in erg s$^{-1}$
\item[6] 6. $T_2$ is the time that jet becomes noticeably unstable, if it does.
\item[7] 7. $S_{j}$ is the distance out to which the jet remains stable, if it goes unstable.
\item[8] 8. $End$ indicates whether or not the bow shock reaches (or goes beyond) the end of the grid before $T_{1}$.
\item[*] $*$ The run was also computed for an overpressure: $P_{j}/P_{a} =10$.
\item[$\dagger$] $\dagger$ The run was also computed with $\Gamma = 4/3$.
\end{tablenotes}

 \begin{figure}
   \begin{minipage}{\textwidth}
   \centering
   \includegraphics[trim=4cm 0cm 4cm 3cm,width=0.32\textwidth,clip]{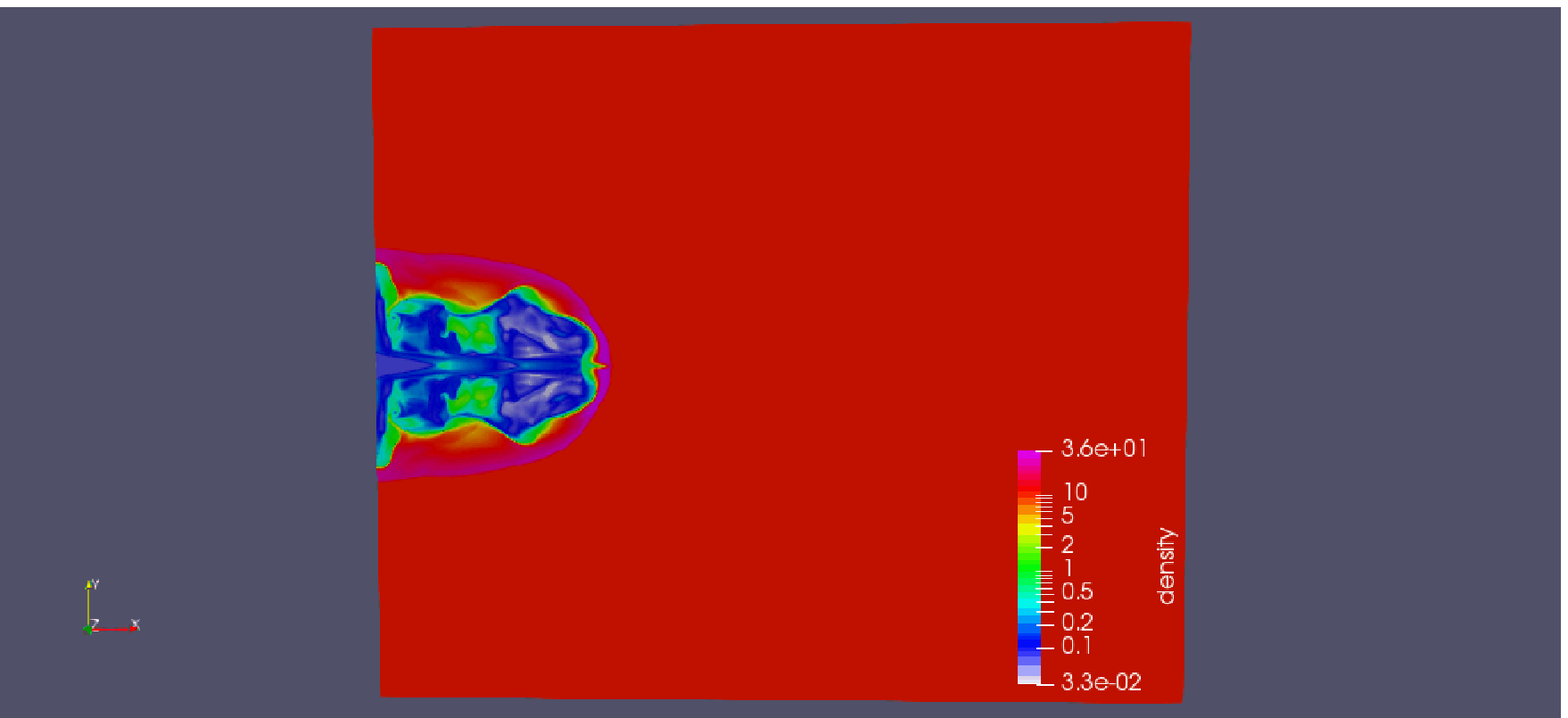}
   \includegraphics[trim=4cm 0cm 4cm 3cm,width=0.32\textwidth,clip]{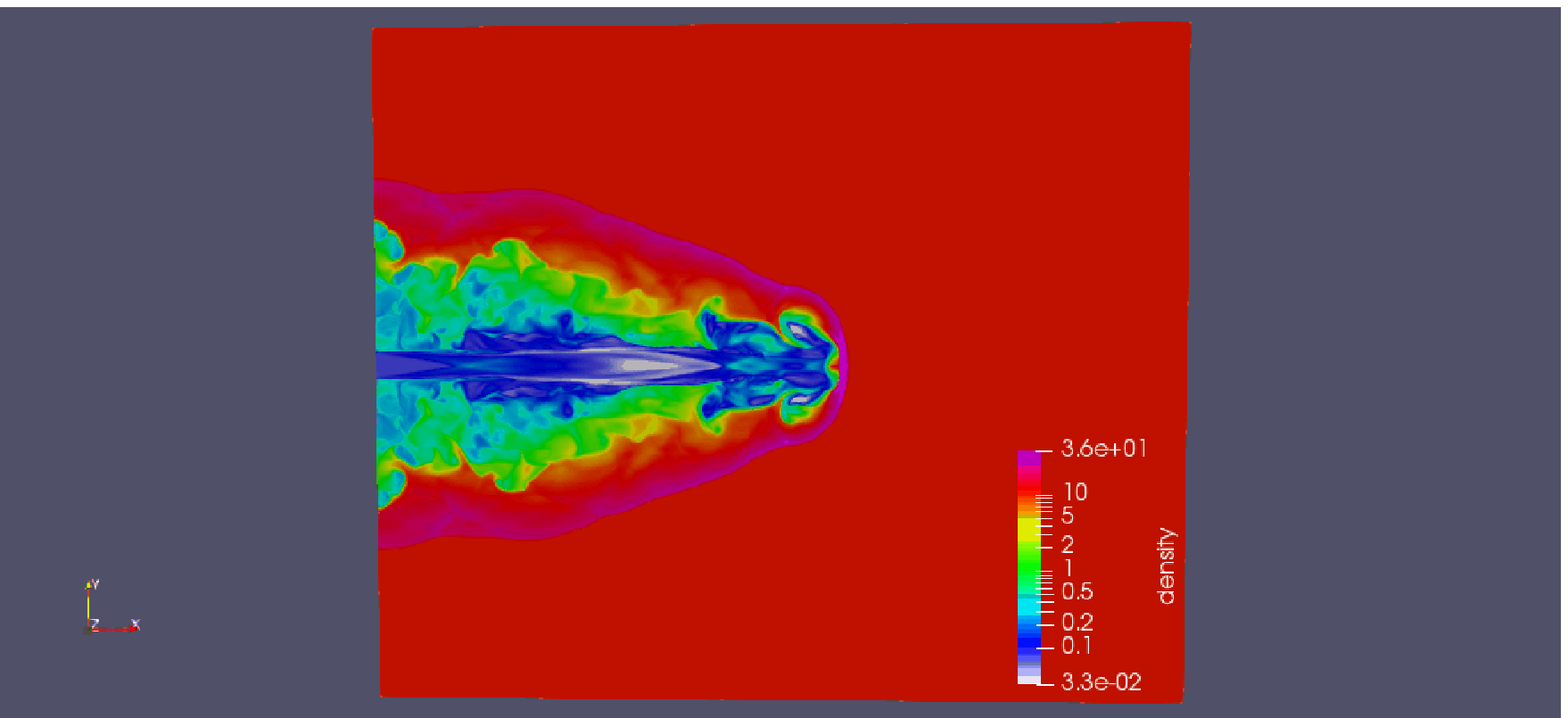}
   \includegraphics[trim=4cm 0cm 4cm 3cm,width=0.32\textwidth,clip]{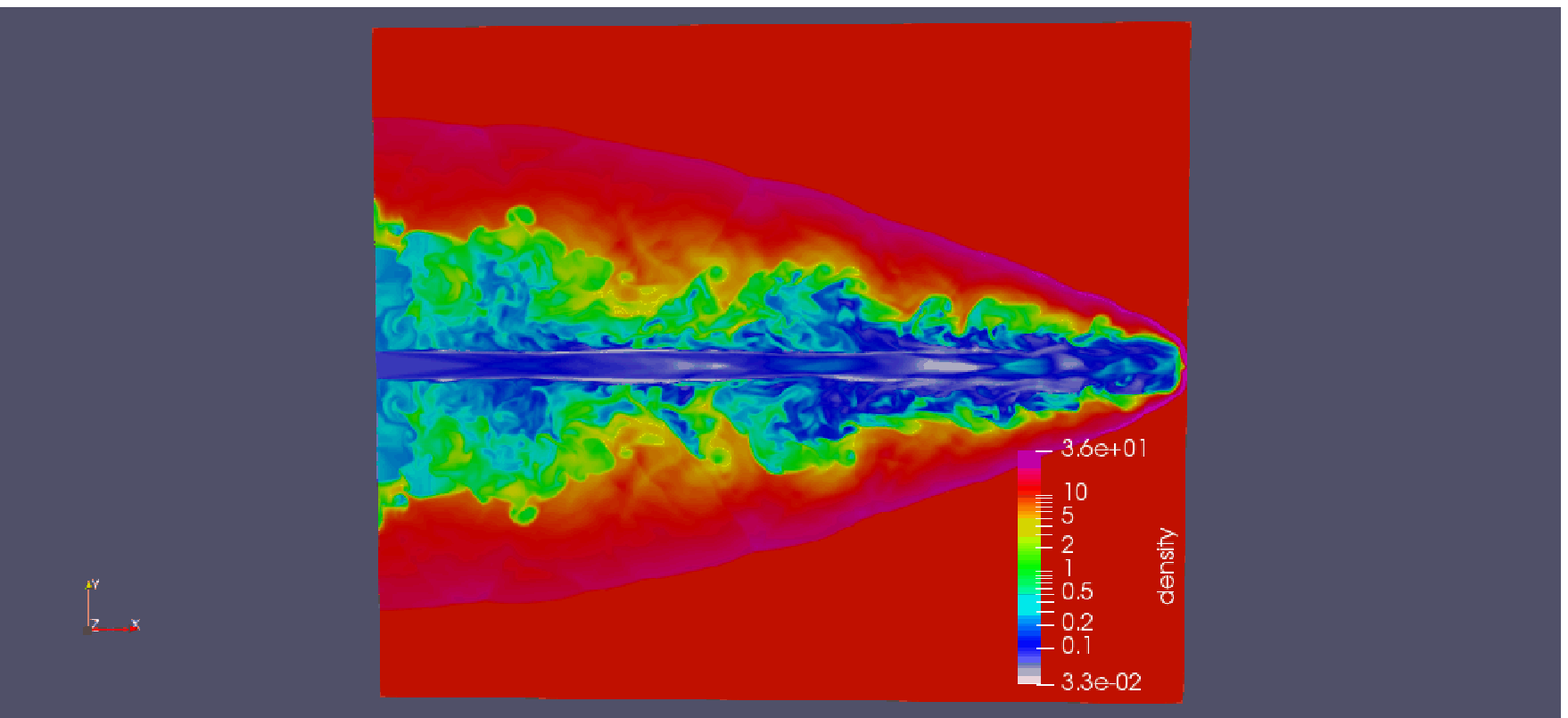}
   \end{minipage}\vspace{0.001cm}
   \caption{Snapshot slices of fluid densities (on a logarithmic scale) for $v_{j} = 0.98c$, $\eta$ = 0.0075, at different stages: (left) $t = 60$; (center) $t=120$; (right) $t=190$. There are 10 zones per jet radius and $600 \times 500 \times 500$ zones to 60 $R_{j}$ along the $x$-axis and to $\pm 25 R_j$ along the $y$- and $z$-axes.}
   \label{fig:3time}
\end{figure}

\section{Results}
 Snapshots of fluid density slices (on a logarithmic scale) in the $z = 0$ plane for a stable jet with $v_{j} = 0.98c$ and $\eta = 0.0075$ at three different time steps are shown in Figure \ref{fig:3time}. All of the ambient densities are taken as fixed and the same (a non-dimensional value of 10.0, denoted by a red color), in our simulations. The length of this simulation, and roughly half of our others,  is 60 $R_{j}$, and the time steps shown are $t =$ 60, 120, and 190. For the three images the scale is such that the maximum density (purple) is in the bow shock and is $\simeq 3.6$ times that of the ambient density (red) and the minimum density (white) is $\simeq 3.3 \times 10^{-3}$ of the ambient density, or a little less than half that of the inflowing jet matter.

 The highest density region is the bow shock advancing into the ambient medium, and behind it is the shocked ambient medium which appears in shades of red, orange and yellow in these log$~\rho$ plots.  A contact discontinuity separates this material from the cocoon of shocked jet material which is left behind as the jet advances and appears in shades of green and blue.   Substantial vortices form within the cocoon and these, along with Kelvin-Helmholtz instabilities, facilitate the mixing of shocked ambient and shocked jet fluids.  A series of strong internal shocks and rarefactions form within the jet, as can be seen by the changes from the blue color corresponding to the injected jet density to the turquoise and white colors, respectively. The first such shock is called the reconfinement shock (even though subsequent ones also help keep the jet confined) and it maintains a relatively constant location after the jet achieves a length some ten times its radius.  In this case, the jet remains well collimated with a terminal shock (corresponding to an observed radio hot spot) until the end of the problem grid is reached. It is thus classified as a stable source, with a cocoon similar in shape to many FR II RGs.   For these simulations we  examine 1D plots of the density or pressure along the jet axis and for the ones we call stable the strongest shock in the jet itself occurs at this termination point, which remains very close behind the bow shock.  At early times the jet is essentially axially symmetrical, but as it propagates outward, small numerical fluctuations grow and modest asymmetries have clearly arisen before the jet leaves the grid.  When we compare simulations with the same input parameters made at moderate resolutions (10 zones across the jet diameter) with our higher resolution runs (20 zones per diameter) we see few significant differences in the distances traveled at given times, the stability, or the gross structure of the shocks; however, the higher resolution runs certainly resolve more features of the internal jet shocks and better display the eddies within the cocoon.

\begin{figure*}
\begin{minipage}{\textwidth}
\centering
\includegraphics[trim=0.7cm 0.5cm 0cm 4cm,width=0.8\textwidth,clip]{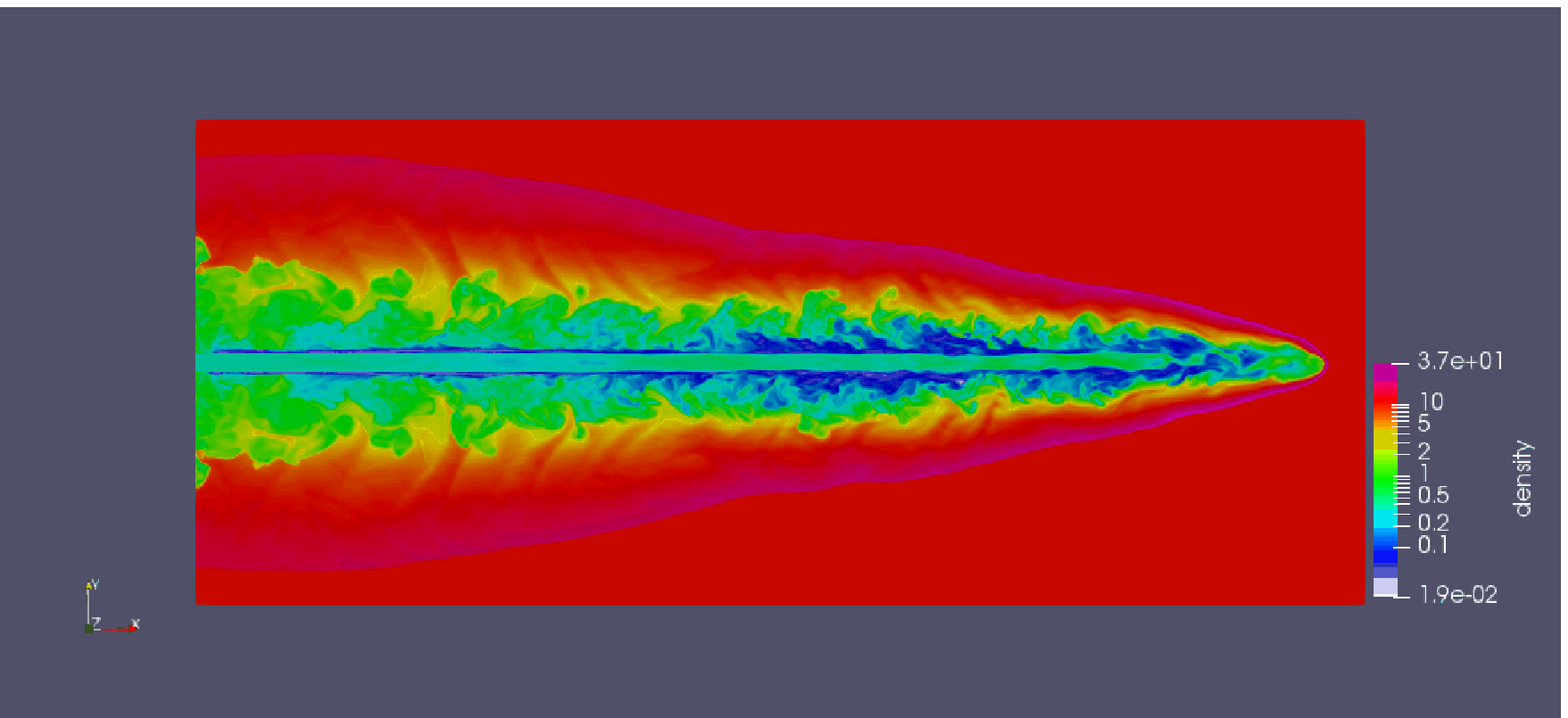}
\includegraphics[trim=0.2cm 0.1cm 0cm 4.4cm,width=0.8\textwidth,clip]{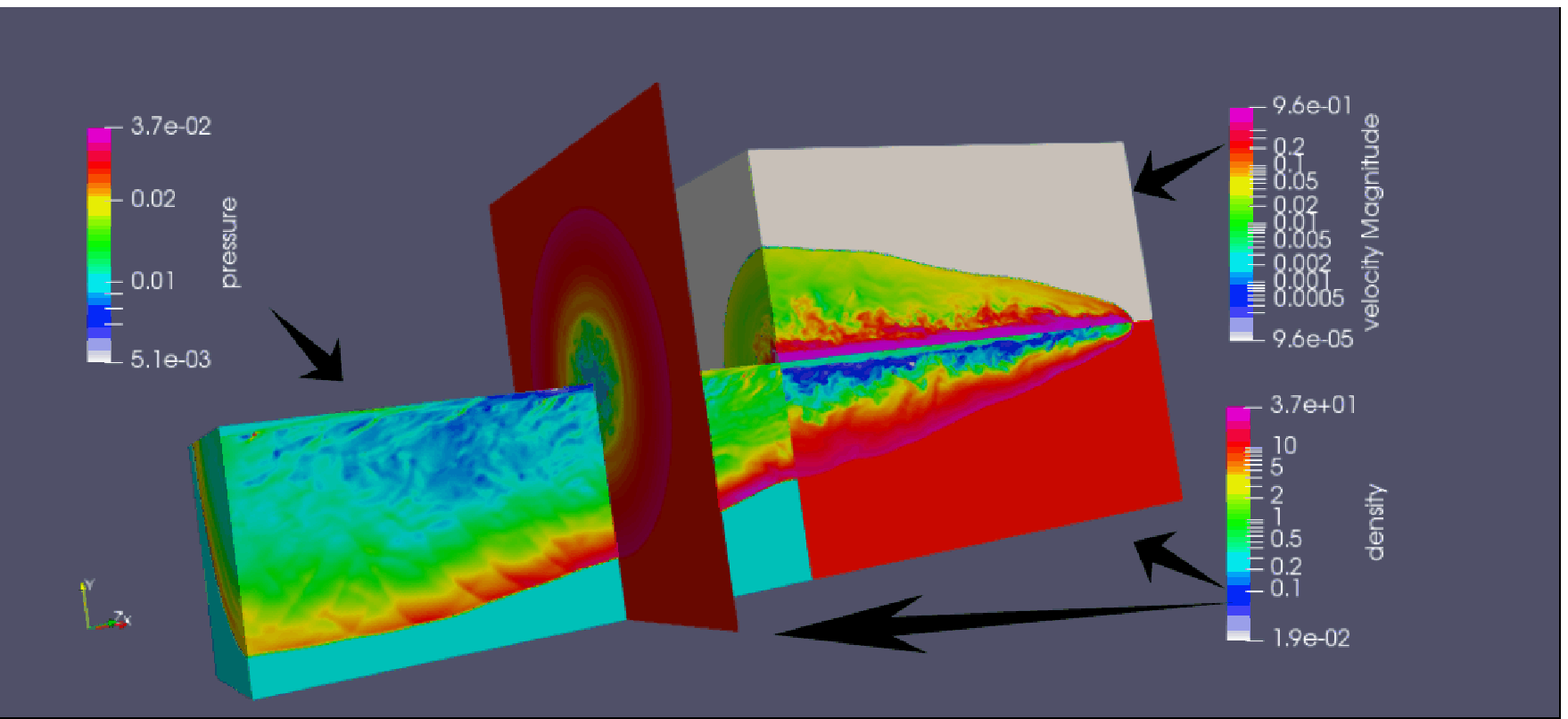}
\end{minipage}\vspace{0.001cm}
\caption{3D jet simulation with parameters $v_{j} =0.96c$ and $\eta$ =0.0316 (high resolution, at $t=250$): (top) longitudinal slice at $z=0$ of the logarithms of the fluid densities; (bottom) pressure (also on a logarithmic scale) shown between $0 < x < 60$, and magnitude of the velocity and density (also on logarithmic scales), shown for $60 \le x < 120$, with velocity shown for $y > 0$ (including a quarter-cross section at $x = 60, y >0, z \ge 0$ and density for $y < 0$; in addition, most of a cross-section of $\rho$ for $x = 37.5$ is shown.}
\label{fig:3Dcross}
\end{figure*}

 Another high resolution simulation of a somewhat slower, but denser, jet (with $v_{j} =0.96c, \eta =0.0316$) is shown in Figure \ref{fig:3Dcross}; in this case the simulation ran to a distance of 120$R_{j}$, and the plots are made shortly before it reaches that distance, at $t = 250$.  The upper panel shows a log$~\rho$ slice similar to that shown earlier for the other jet in Figure \ref{fig:3time}.  The lower-panel of Figure \ref{fig:3Dcross} presents different  cross-sections of the pressure, density and velocity distributions. The top plot shows that the jet in the center remains clearly  stable for over 70\% of its length, but it starts to wiggle a bit near the end of this 120 $R_{j}$ grid. This means it may eventually transition into an  FR I morphology at an even later time, though it might also appear as an FR II with hot-spot offset from the current jet direction; such morphologies are common, including in the canonical FR II, Cygnus A \citep[e.g.][]{Carilli96}. The left portion of the bottom combined plot illustrates the typical pressure structures for these strong, pressure-matched, jets, where both the ambient and initial jet pressures show as a light blue shade while a much higher pressure (in red) is seen in the bow shock and much of the shocked ambient gas (red to yellow).  The shocked  material that passed through the jet and fills the radio emitting cocoon contains only modest pressure variations (shades of blue) and does not differ greatly from the average internal jet pressure.  Within the jet itself, pressure variations corresponding to shocks and rarefactions are visible.  The logarithm of the magnitude of the velocity is given in the upper-right portion of the lower panel of Figure \ref{fig:3Dcross}.   The unshocked ambient medium, which doesn't move, is shown as white and the shocked ambient medium, which is moving very slowly ($< 0.05 c$) appears as yellow and green. The cocoon moves at modest speeds (up to $\sim 0.2 c$), but the only strongly relativistic motions are seen within the jet itself, where the average velocity stays very close to the injection speed of $0.96 c$ (purple). The cross section of density shows a nearly axisymmetric distribution, particularly in the shocked ambient material, but  modest perturbations have appeared within the cocoon material; the apparent dominance of an $m=4$ mode at this stage is almost certainly attributable to the use of cartesian coordinates to model a cylindrical jet.

\begin{figure*}[!htbp]
\begin{minipage}{\textwidth}
\centering
\includegraphics[trim=4cm 0cm 2cm 4.5cm,width=0.45\textwidth,clip]{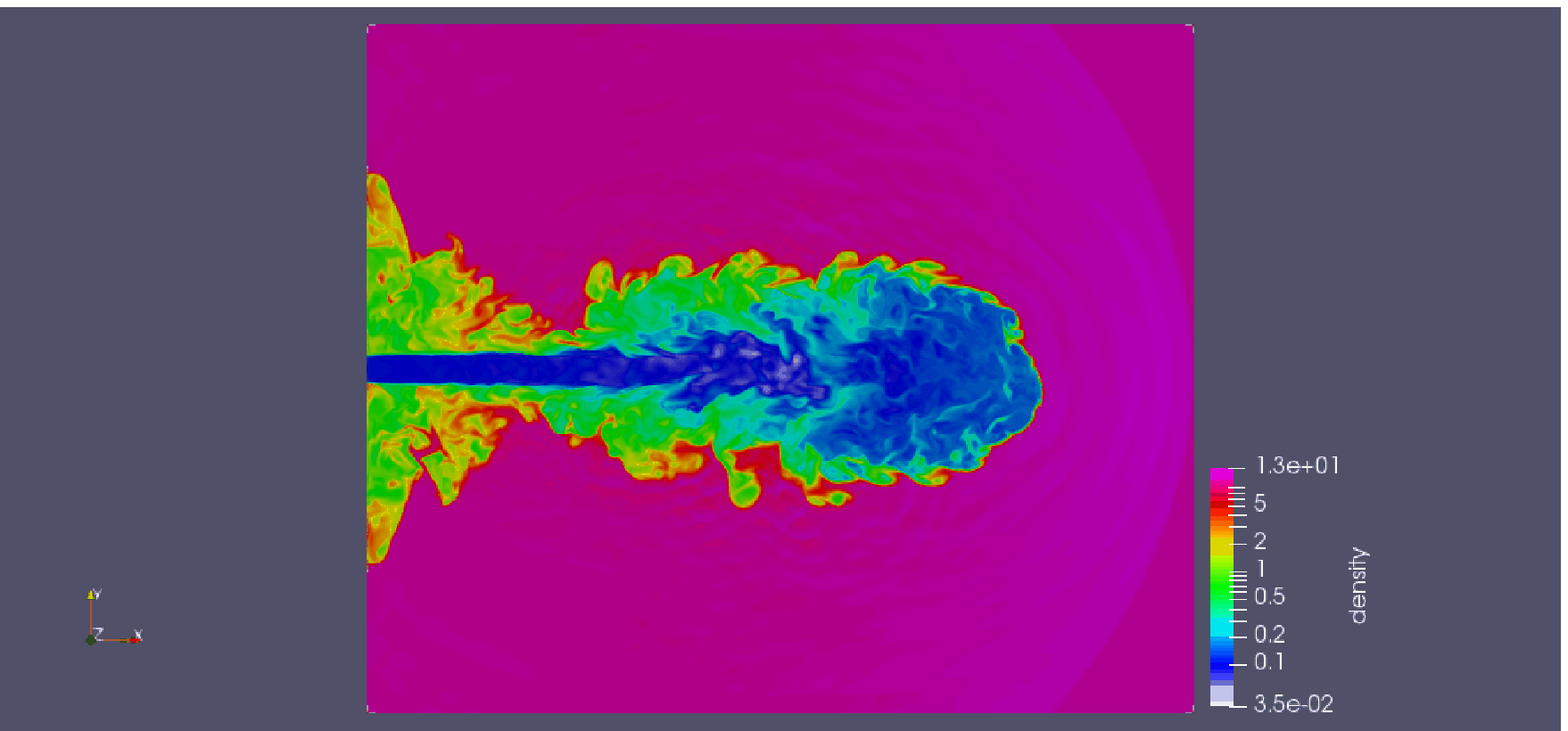}
\includegraphics[trim=4cm 0cm 2cm 4.5cm,width=0.45\textwidth,clip]{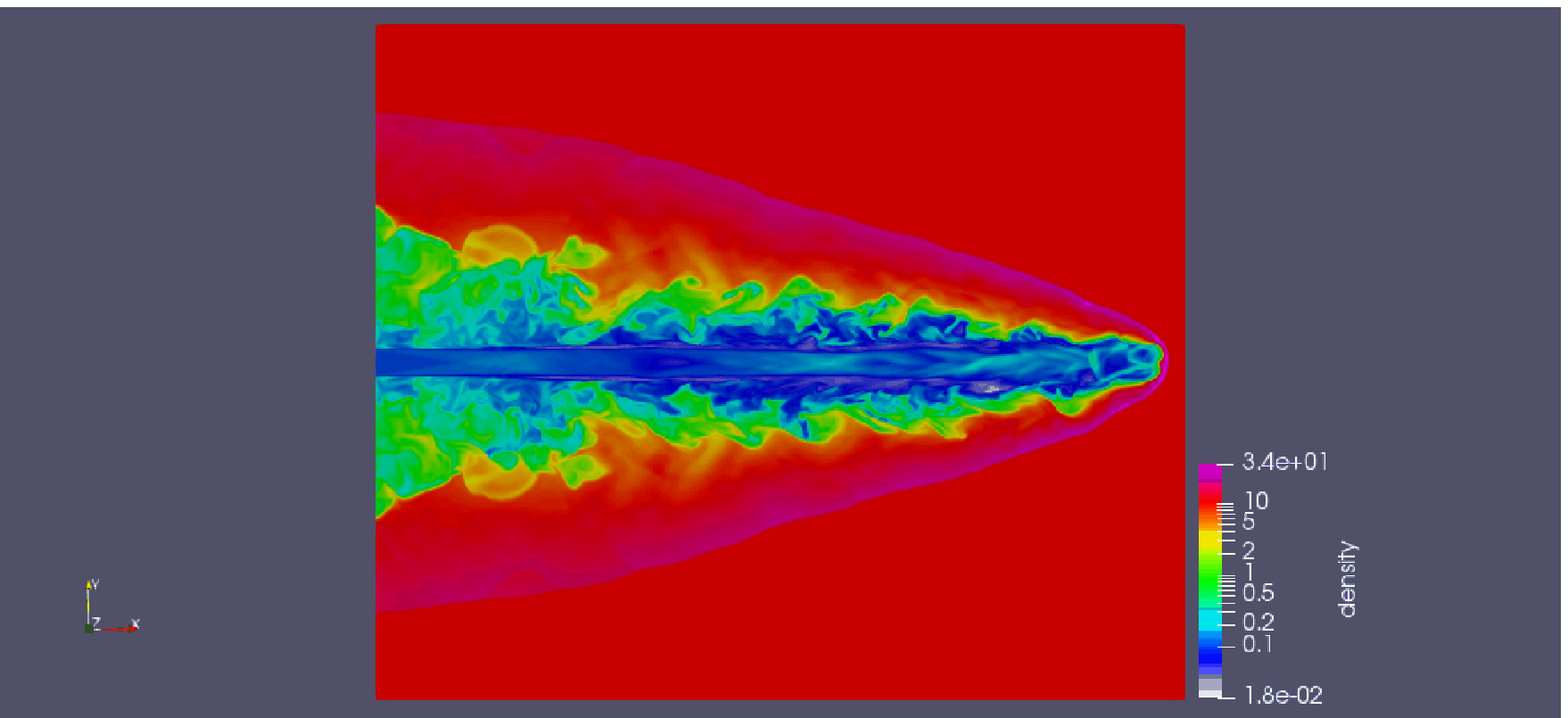}
\end{minipage}\vspace{0.12cm}
\begin{minipage}{\textwidth}
\centering
\includegraphics[trim=4cm 0cm 2cm 4.5cm,width=0.45\textwidth,clip]{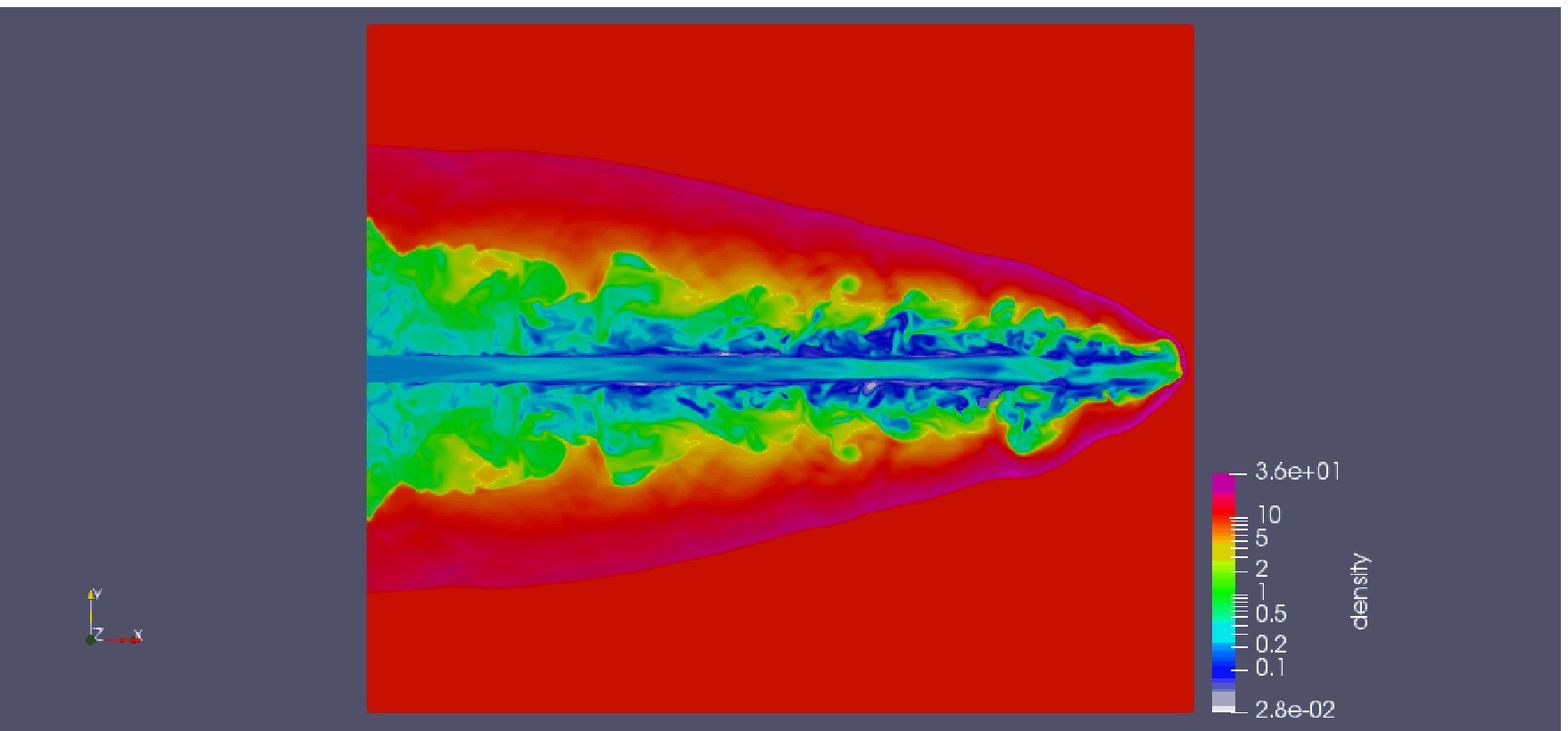}
\includegraphics[trim=4cm 0cm 2cm 4.5cm,width=0.45\textwidth,clip]{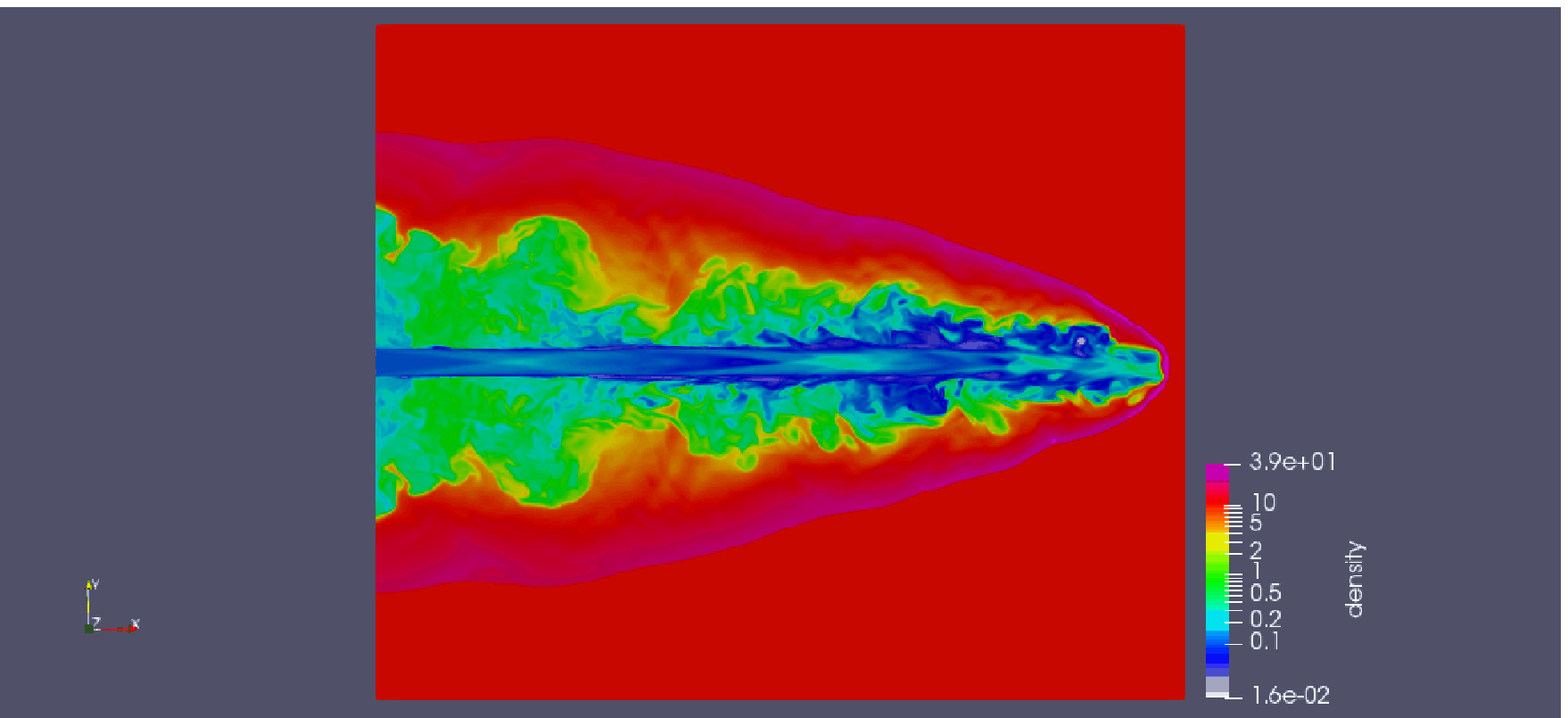}
\end{minipage}
\begin{minipage}{\textwidth}
\centering
\includegraphics[trim=4cm 0cm 2cm 4.5cm,width=0.45\textwidth,clip]{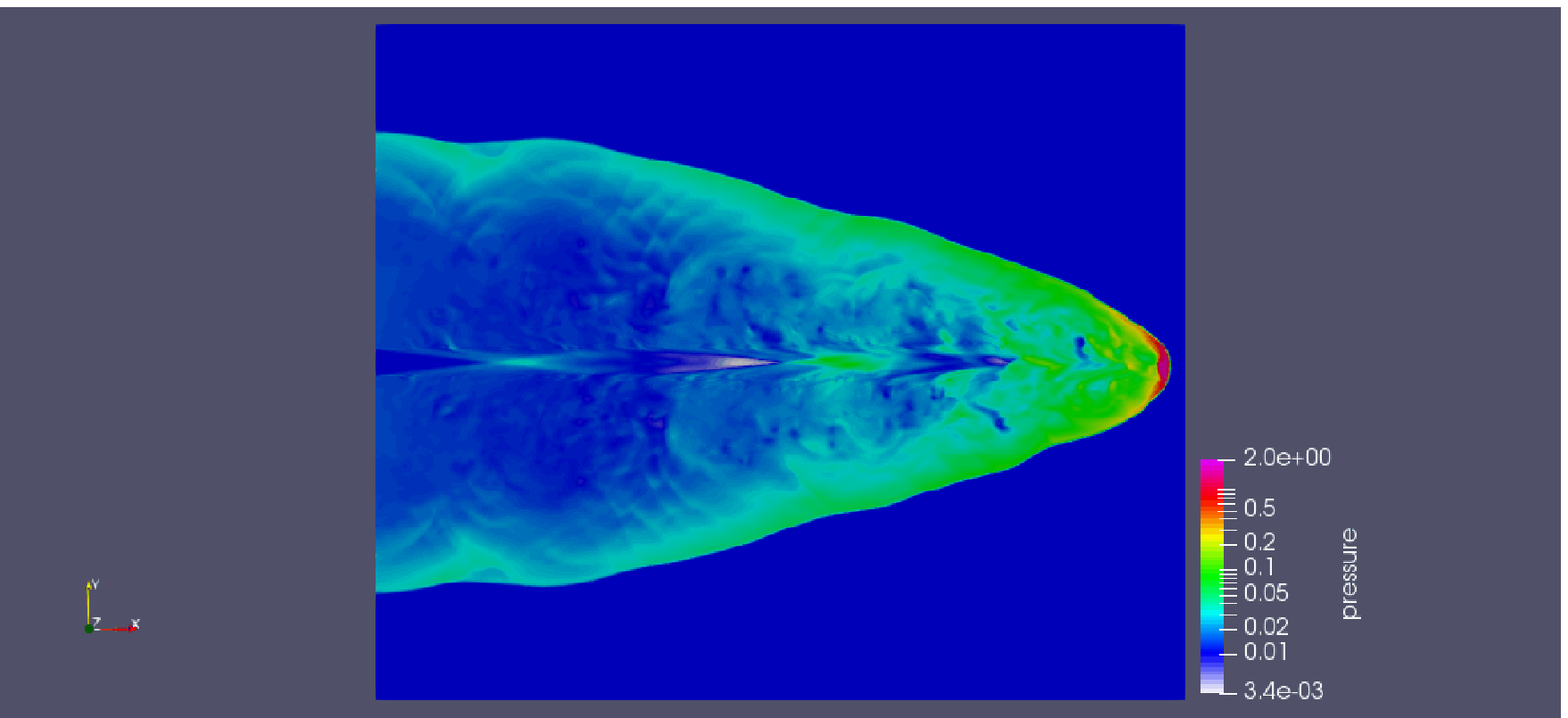}
\includegraphics[trim=4cm 0cm 2cm 4.5cm,width=0.45\textwidth,clip]{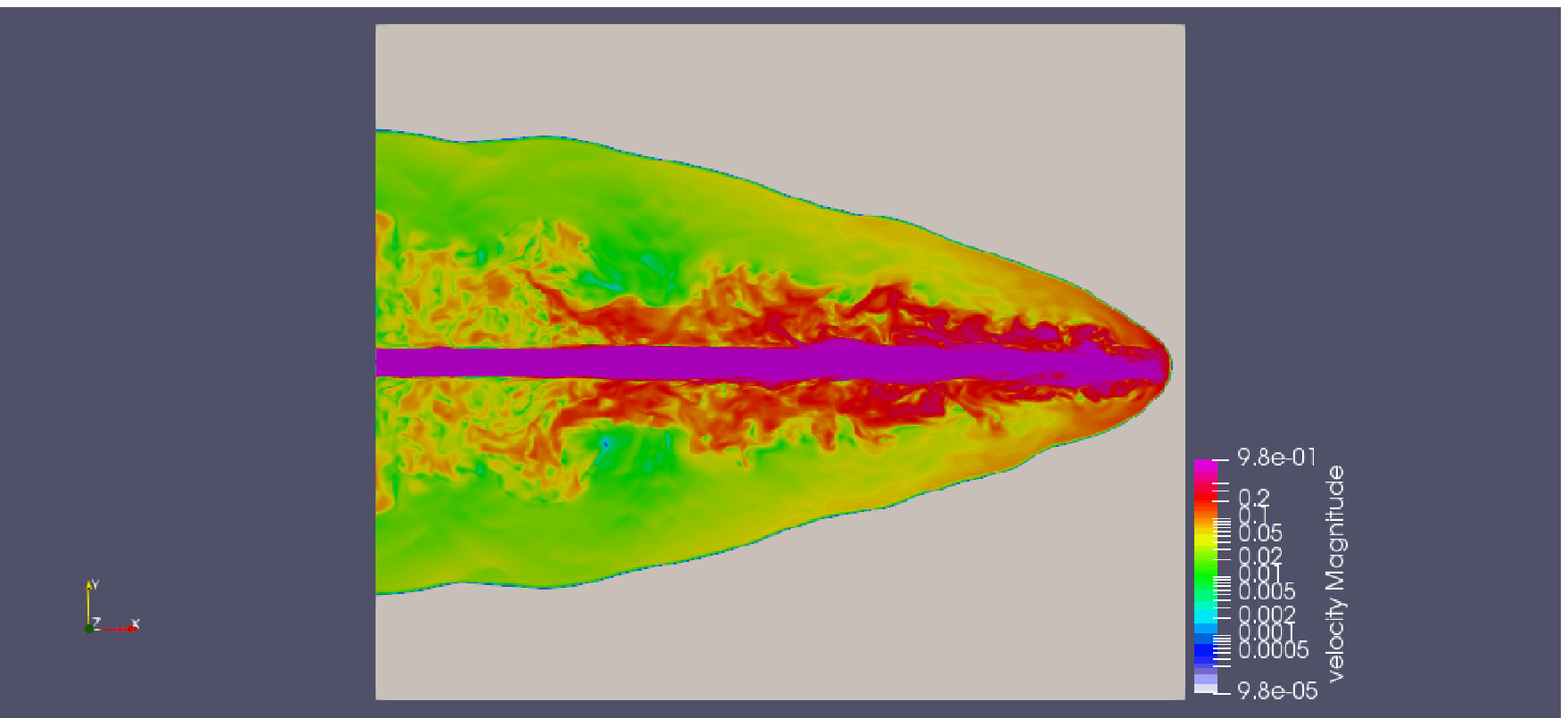}
\end{minipage}
\caption{High resolution simulations shown at or very close to 60 $R_{j}$: (top left)  $\rho$ for $v_{j} =0.70c, \eta =0.01$ at $t =970$;
(top right)  $v_{j} =0.96c$, $\eta =0.01$ at $t=208$; (middle left) $\rho$ for $v_{j} =0.96c, \eta =0.02$ at $t=170$; (middle right) $\rho$ for $v_{j} =0.98c, \eta =0.01$ at $t=168$; (bottom left) $P$ for $v_{j} =0.98c, \eta =0.01$ at $t=168$;  (bottom right) $|v|$ for $v_{j} =0.98c, \eta =0.01$, also at $ t=168$.}
\label{fig:Highruns}
\end{figure*}

Four different high resolution simulations extending out to 60 $R_{j}$ are shown in Figure \ref{fig:Highruns}. The top four plots are the snapshots of slices of the fluid densities (on a logarithmic scale) with the input parameters:  $v_{j} =0.70c$, $\eta =0.01$ (top left); $v_{j} =0.96c$, $\eta =0.01$ (top right); $v_{j} =0.96c$, $\eta =0.02$ (middle left); and $v_{j} =0.98c$, $\eta =0.01$ (middle right). The bottom panels give the distribution of pressure (bottom left) and velocity (bottom right) for that last simulation. There are clearly different morphology evolutions displayed.  For $v_{j} =0.70c$, $\eta =0.01$, the jet starts to become unstable at the distance of 24.0 $R_{j}$ which has been listed in Table \ref{tab:parameter} and as it continues to develop it would certainly produce an  FR I morphology on extragalactic scales, particularly when ambient material is entrained in the jet and renders it transonic \citep[e.g.][]{Bick94,Bick95}.    We define the position at which the jet terminates within the lobe as the strongest density enhancement measured along the jet axis.  In these unstable simulations, any additional density enhancements further out toward the hotspots are weaker and visual inspection of 2D density plots shows they involve  swinging or effectively fragmented flows.  In such relatively weak jets the bow shock actually can fade into a subsonic pressure wave as is seen in the top left panel of Fig.\ \ref{fig:Highruns}. The second case, with $v_{j} =0.96c$ and $\eta =0.01$, is essentially stable out to this distance,  with the strongest jet shock still close behind the bow shock, but some wiggles are developing in the jet and it is possible that it will become unstable shortly. The other two simulations remain fully stable while propagating past the end of the 60 $R_{j}$ grid and would exhibit FR II morphology; these differences will be discussed further in Section 5.1.

The pressure distribution at the  bottom left of Figure \ref{fig:Highruns} shows that the pressure at the head of bow shock is much higher than that in the jet material but a high pressure terminal shock is also present at the end of the jet just behind the bow shock, along with three internal shocks, which would likely be viewed as knots within the jet. The velocity structure shown in the lower right panel makes it clear that the jet velocity remains highly relativistic until the terminal shock and that the flow in the cocoon is much slower.  Hence, even though the cocoon volume is much higher than that of the jet, if the source is viewed close to the line of sight the much greater Doppler boosting of the jet would mean that emission from the jet would dominate the observed brightness.

\begin{figure*}
\centering
\includegraphics[trim=0.5cm 0.2cm 0cm 4.4cm,width=0.85\textwidth,clip]{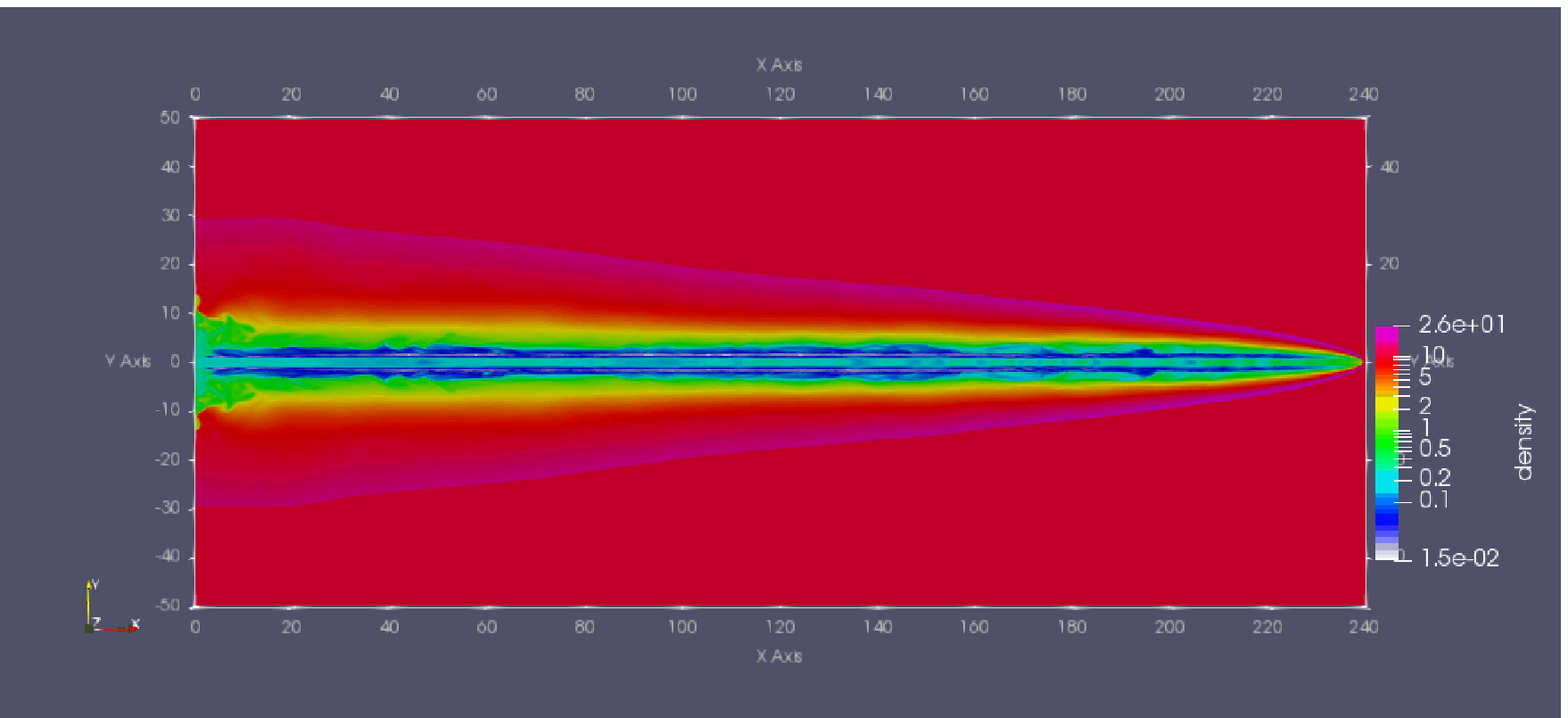}
\caption{Density slice for a simulation extending to 240 $R_{j}$ performed at moderate resolution for input parameters of $v_{j} =0.985c, \eta =0.02$ at $t=369$.}
\label{fig:long240}
\end{figure*}

An even more powerful jet, with $v_j=0.985c, \eta =0.02$, is shown to remain beautifully stable all the way out to 240 $R_{j}$ in Figure \ref{fig:long240}.  At this high power the cocoon is very narrow and it would not resemble the typical FR II extragalactic radio source morphology.  However, if scaled down to galactic nuclear scales it is reminiscent of the morphology of many VLBI jets \citep[e.g.][]{Marti95,Marti97,Miod97}.

\section{Estimation of Emission Brightness}

 Any radiating material traveling at relativistic speed will be affected by Doppler boosting, which strongly beams the emission in the direction of its velocity, providing an amplification in brightness and a shortening of the observed time over which a variation occurs \citep[e.g.][]{Urry95,Gopal03}. The flow in the entire jet wiggles back and forth by a small amount and there are certainly differential motions within the jet.  Hence, the angle of the velocity of each cell with respect to the observer is always changing, resulting in the apparent rise and fall of the received flux from individual cells within the jet region.  In addition, the density and pressure of each cell varies with time.  The Doppler factor for a cell characterized by $(x,y,z)$ coordinates and numerated by ${i,j,k}$ is given by
  \begin{equation}
   \delta_{ijk} = \frac{1}{\gamma_{ijk} (1-\beta_{ijk} \cos\theta_{ijk})},
  \label{equ:doppler}
  \end{equation}
 where $\gamma_{ijk}$ is the Lorentz factor of the flow in that voxel, $\beta_{ijk}$ is its speed in units of $c$, and $\theta_{ijk}$ is the angle of the velocity of that cell with respect to the observer.
 Within our simulations we have the vector velocity and values of density and pressure in every cell and for an assumed value of $\theta$ for the line-of-sight angle to the jet axis we can find $\delta_{ijk}$.

\begin{figure*}
\centering
\includegraphics[trim=0cm 0cm 0cm 0cm,width=0.5\textwidth,clip]{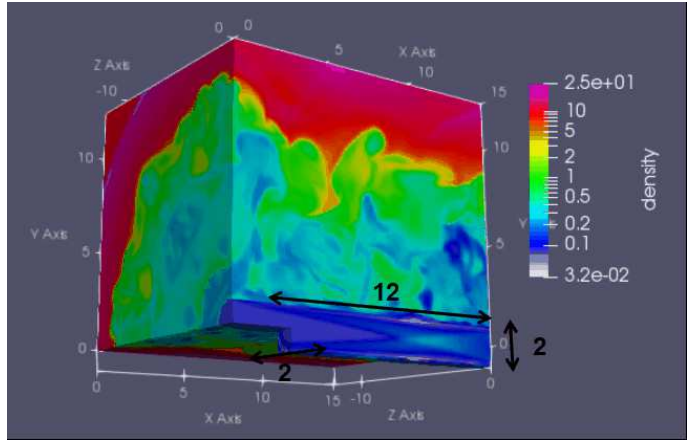}
\caption{The selected emission region, chosen to be close to the reconfinement shock. $v_{j}$=0.99c, $\eta$=0.0075, t=130.}
\label{fig:region}
\end{figure*}

 In principle it  would be optimal to include boosted emissivity from all zones in the jet and cocoon, but this is computationally overwhelming as it must be done for thousands of time intervals in post-processing to produce a light curve.  The regions we used for computing the source brightness were determined visually after examining most of the the entire suite of simulations, since only zones moving at high velocities will obtain strong boosting and only zones with higher densities and pressures are expected to have high rest-frame emissivities.   A location close to the (first) reconfinement shock is optimal \citep{Poll16} and is illustrated in Figure \ref{fig:region}. The explicit range in which we chose to compute emission is 12 $R_{j}$ in length, and 2 $R_{j}$  in width and height ($3 < x \le 15, -1 < y \le +1, -1 < z \le +1$).  Therefore, the total number of zones used for computing the emission are 6000  in medium resolution runs (5 zones in each length unit) and 48000  for high resolution runs (10 zones in each unit).  The input parameters of the example shown in Figure \ref{fig:region} are $v_{j}=0.99c$, $\eta$=0.0075, at $t=130$  (the density for moderate resolution simulation is illustrated). We remove much, but not all, of the initial transient burst in emission produced by the formation of the jet and the passage of the bow shock by only considering light-curves to begin once the recollimation shock had formed and had a nearly stable location (taken to be from $t=150$ in our simulations) and we end the computations once the simulation was recomputed with the dimensionless time extended to 1000 or more units.  Doing so meant that the jets usually propagated beyond the right-hand boundary of the grid, but any loss of material from the grid would have essentially no impact on variations occurring in the inner quarter or eighth of the entire simulation.  We exported data at intervals of 0.25 time-units, thus typically producing 3400 output sets of velocity, density and pressure for the relevant zones between $150 \le t \le 1000$, which was sufficient for use in the calculation of power spectra spanning about three decades in frequency.

The basic emitted flux per unit volume can be considered arbitrary for our purposes as we are only interested in the relative variations, so it was set to unity for an individual zone.  The physical mechanism producing the jet emission from the radio through (at least) soft X-ray bands  is synchrotron emission, the intensity of which scales with the number of relativistic particles and the square of the strength of the magnetic field.  As we have not incorporated the microphysics necessary to compute the  fractions of ultrarelativistic electrons in each zone and  their distributions in energy, we assume that the zone rest mass density can be taken to be proportional to the number of relativistic electrons that produce synchrotron emission in that zone. We also scaled the emission by the pressure in that zone, as the pressure, $P$, can be reasonably taken to be proportional to the square of the magnetic field strength, though of course we have no information on the orientation of that ``field'', so we can say nothing about the polarization of the radiation.   For each zone at each output temporal step, we use the velocity data to calculate a Doppler boosting factor based upon an assumed viewing angle to the $x$-direction along the initial jet direction. The resulting observed flux was then summed along with the observed fluxes of the other zones within the range from that time-step to yield a total observed estimated flux for that time.  We do not take into account the temporal lags for receipt of emission from different zones in the jet as has been done in models or simulations that involve many fewer zones \citep[e.g.][]{Mars14,Calafut15,Poll16}, as this would demand much greater computational resources and is not expected to produce significant differences to the magnitude of amplitude fluctuations nor to the shape of the power-spectra.   With these approximations, the observed flux from this key portion of the jet is estimated as
\begin{equation}
   B = \sum_{ijk} \delta_{ijk}^{m+s} \cdot P_{ijk} \cdot \rho _{ijk} ,
  \label{equ:doppler}
  \end{equation}
 where $\delta_{ijk}$ is the Doppler boosting factor, $P_{ijk}$ is the pressure in each zone, and $\rho_{ijk}$ is the density.
 The observed flux is proportional to the ($m + s$) power of the Doppler boosting factor, where $s$ is the slope of the synchrotron spectrum (with $S_{\nu} \propto \nu ^{-s}$) and $m$ is 2 for continuous flow and 3 for a shock in a flow \citep{Begel84}. We took $s$ to be equal to a typical value of 0.5 for a relativistic jet, and $m$ was set to 2 in this work.

\section{Discussion}

We have carried out a series of  simulations to explore the effects of the different parameters on the jet propagation, evolution, and resulting morphologies. As shown in Section 3, snapshots of density, pressure and velocity retrain almost axisymmetric distributions for some time but eventually numerical perturbations seed a range of instabilities even without any prescribed variations in jet power or direction.   For weaker sources these instabilities basically disrupt the jet and while the bow shock continues to advance without a coherent jet making it to the outer part of the lobe and this can produce a  FR I type morphology.  Most of our simulations were sufficiently powerful so that they indicated essentially stable jets which retain a strong terminal jet shock for extended times; these look like FR II radio sources, particularly for the lower values of $\eta$. We also computed a few simulations that were  over-pressured, with $P_{j} / P_{amb} = 10$ and in this section we also  consider the differences in morphology  between 3D and 2D (slab-like) simulations.

\subsection{Distinction between FR I and FR II types}

\begin{figure}
    \centering
    \includegraphics [trim=0.2cm 0cm 0.2cm 0cm,width=0.65\textwidth,clip]{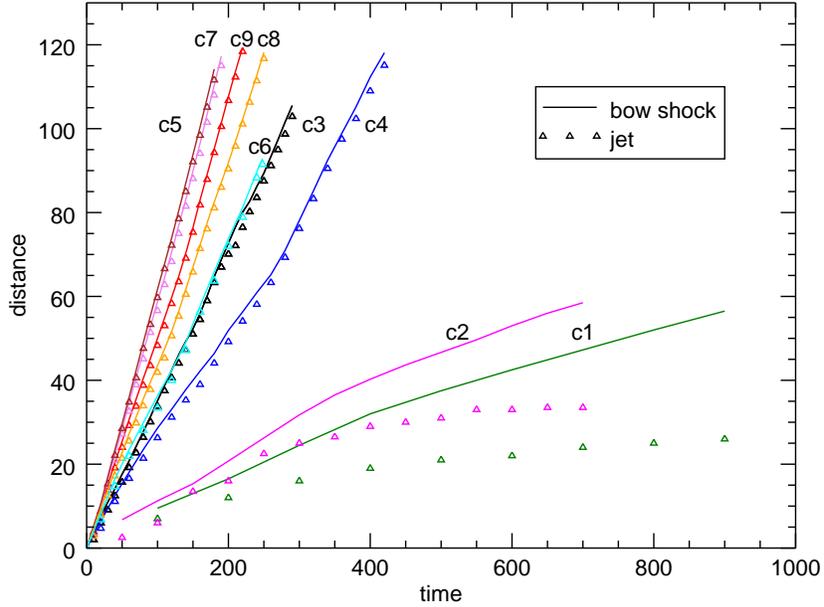}
    \caption{Solid curves are the leading edges of the bow shocks, and the triangles give the jet termini behind the bow shocks for 9 simulations,  labeled c1 to c9; parameters for these runs are  listed in Table \ref{tab:runs}.}
    \label{fig:distance}
 \end{figure}

Although our focus has been on the propagation of powerful, stable jets, we did consider a range of jet powers that spanned a factor of 985 in order to study the distinction between FR I and FR II morphologies.  To illustrate some of the key differences, plots of the distances against time of both the bow shock  position along the x-axis (or peak of the wave front separating uncompressed ambient medium from that affected by the jet) and jet terminus (the strongest shock within the jet) for nine different runs have been shown in Figure \ref{fig:distance}. These are plotted with different colors for different input parameters, marked as c1 -- c9 on the figure, with properties listed in Table \ref{tab:runs}. Solid lines are the spread of bow shocks and triangles are jet termini, (as defined in Sect.\ 3) which always start out close to the bow shock.  The first two belong to FR I morphologies and their slower growth puts them on the right side of Figure \ref{fig:distance}, and the others are FR II, as shown on the left 7 curves.

   \begin{table}
    \centering
    \caption{Parameters for bow-shock and jet terminus distance  plots}
    \label{tab:runs}
    \begin{tabular}{llllllllll}
    \hline\hline
         Run     &  c1  &  c2  &  c3   &  c4   &  c5   &  c6   &  c7   &  c8   &  c9  \\\hline
      $\beta_{j}$  & 0.7  & 0.8  & 0.96  & 0.97  & 0.98  & 0.985 & 0.985 &  0.99 & 0.993 \\
      $\gamma_{j}$ & 1.40 & 1.67 & 3.57  & 4.11  & 5.03  & 5.79  & 5.79  &  7.09 & 8.47  \\
      $\eta$       & 0.01 & 0.01 & 0.02  &0.0075 &0.0316 & 0.02  & 0.0075& 0.0075&0.0075 \\
      $L_j~ \rm{(erg ~s}^{-1})$	  &8.05(44)&1.82(45)&3.58(46)&1.90(46)&1.27(47)&1.11(47)&4.18(46)&8.70(46)&9.58(46)\\
      Type     & FR I & FR I & FR II & FR II & FR II & FR II & FR II & FR II & FR II \\
      \hline
   \end{tabular}
  \end{table}

\begin{figure*}[!htbp]
\begin{minipage}{\textwidth}
\centering
\includegraphics[trim=4cm 0cm 2cm 4.4cm,width=0.45\textwidth,clip]{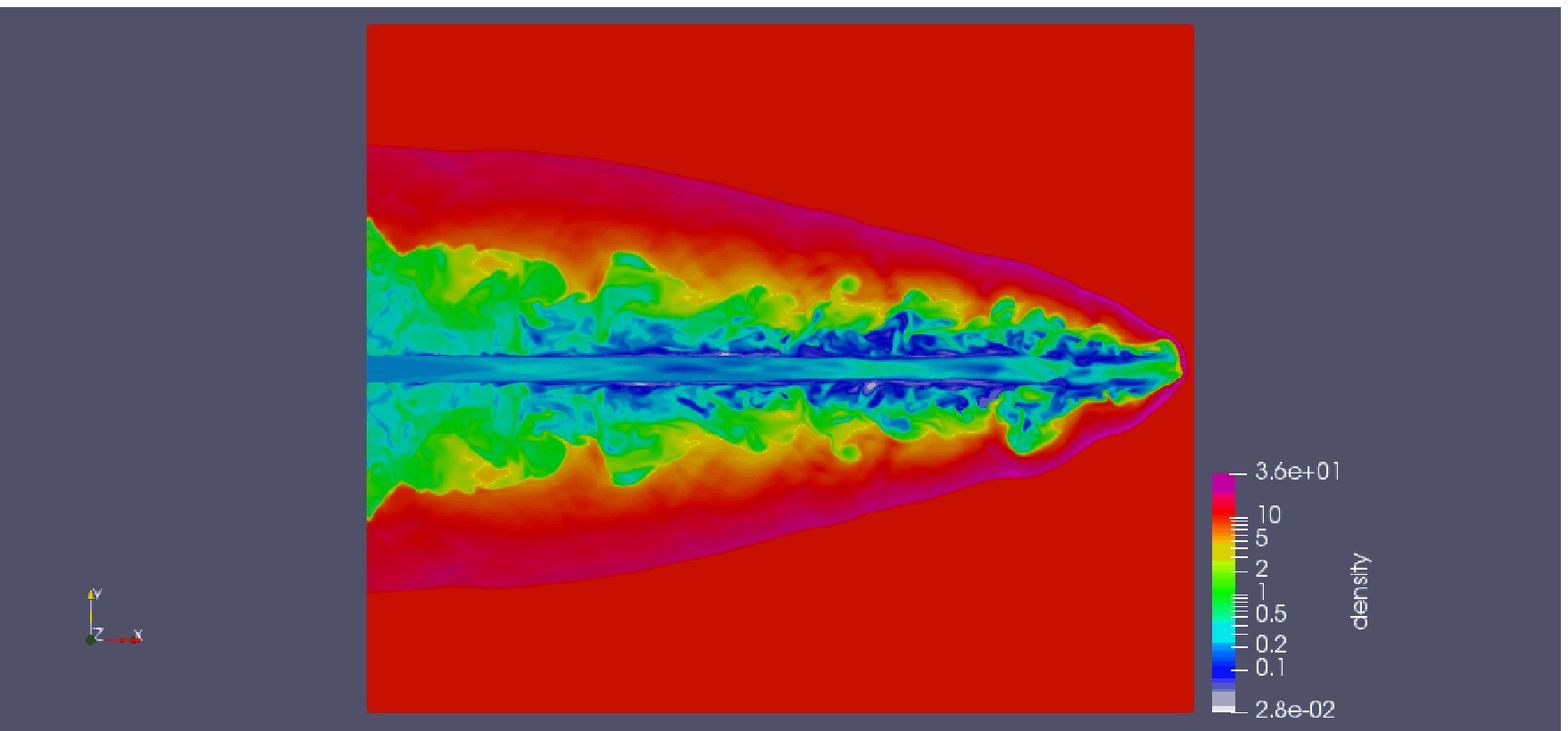}
\includegraphics[trim=4cm 0cm 2cm 4.4cm,width=0.45\textwidth,clip]{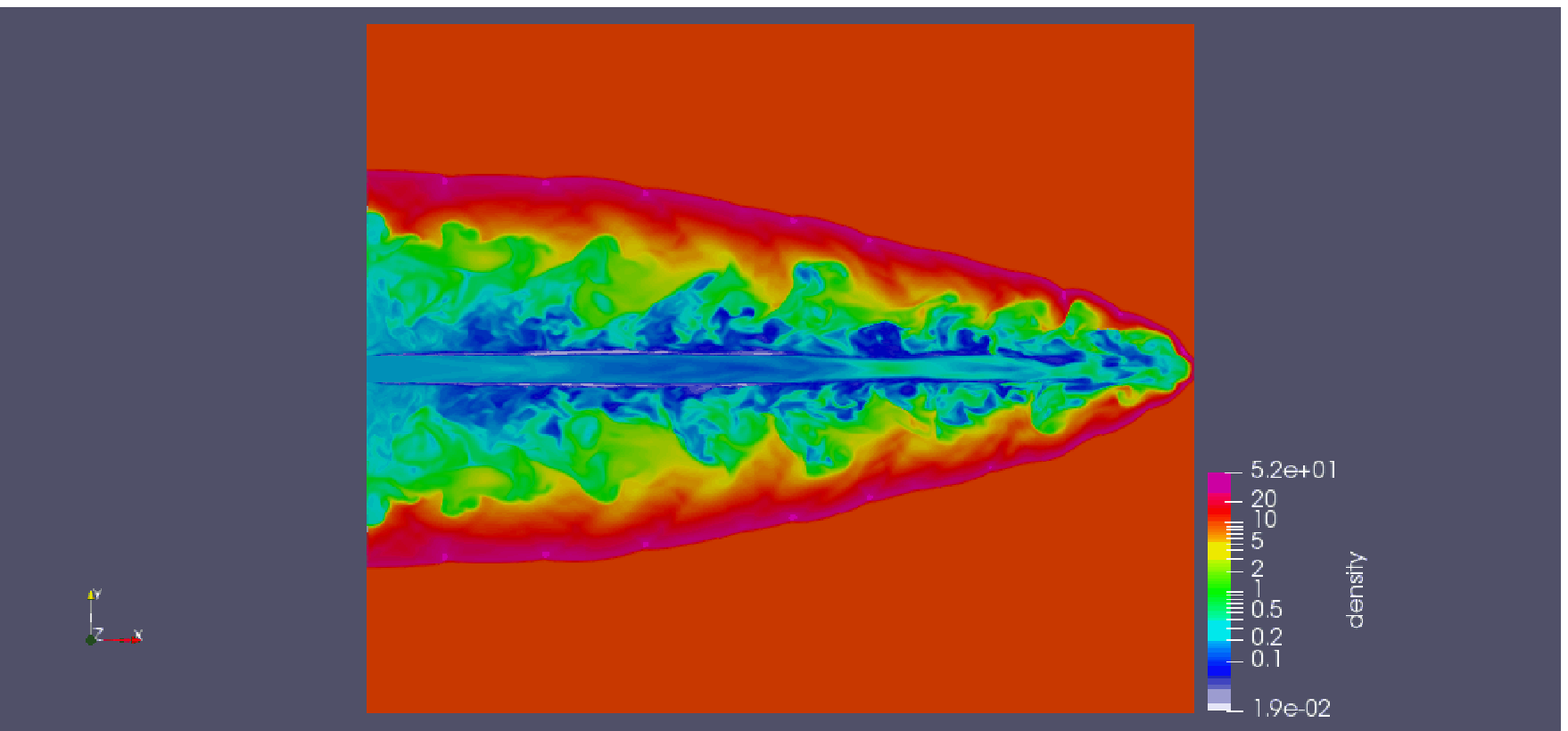}
\end{minipage}\vspace{0.1cm}
\begin{minipage}{\textwidth}
\centering
\includegraphics[trim=4cm 0cm 2cm 4.4cm,width=0.45\textwidth,clip]{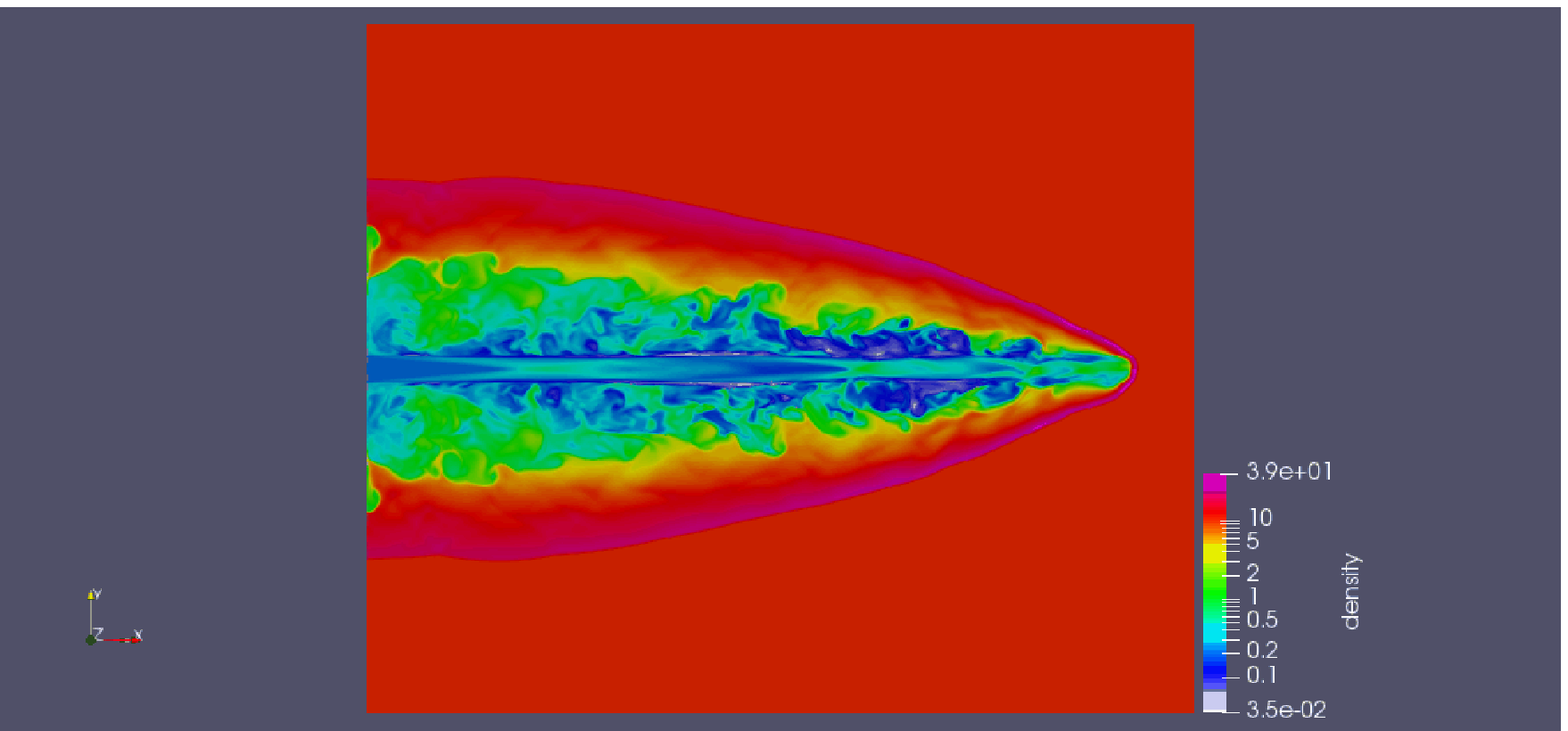}
\includegraphics[trim=4cm 0cm 2cm 4.4cm,width=0.45\textwidth,clip]{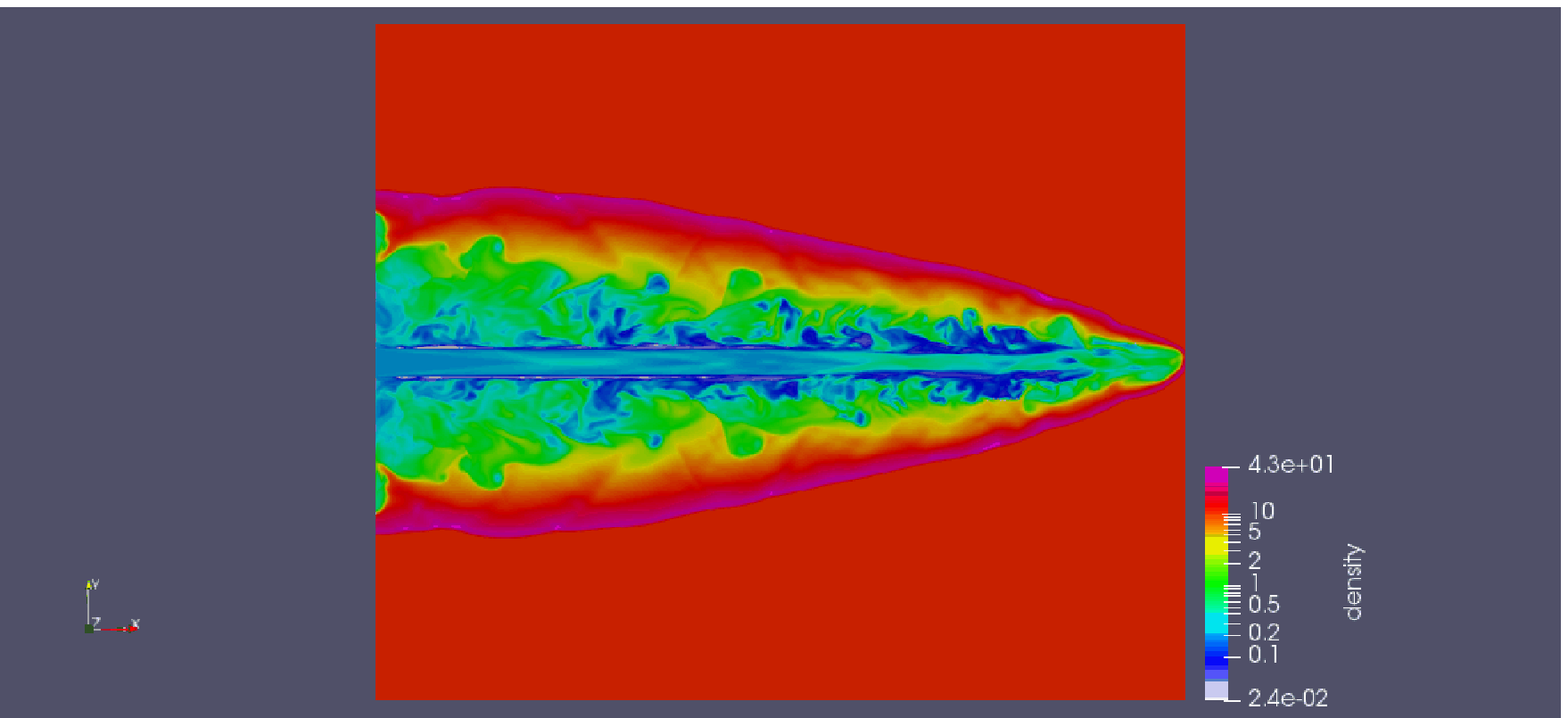}
\end{minipage}\vspace{0.1cm}
\begin{minipage}{\textwidth}
\centering
\includegraphics[trim=4cm 0cm 2cm 4.4cm,width=0.45\textwidth,clip]{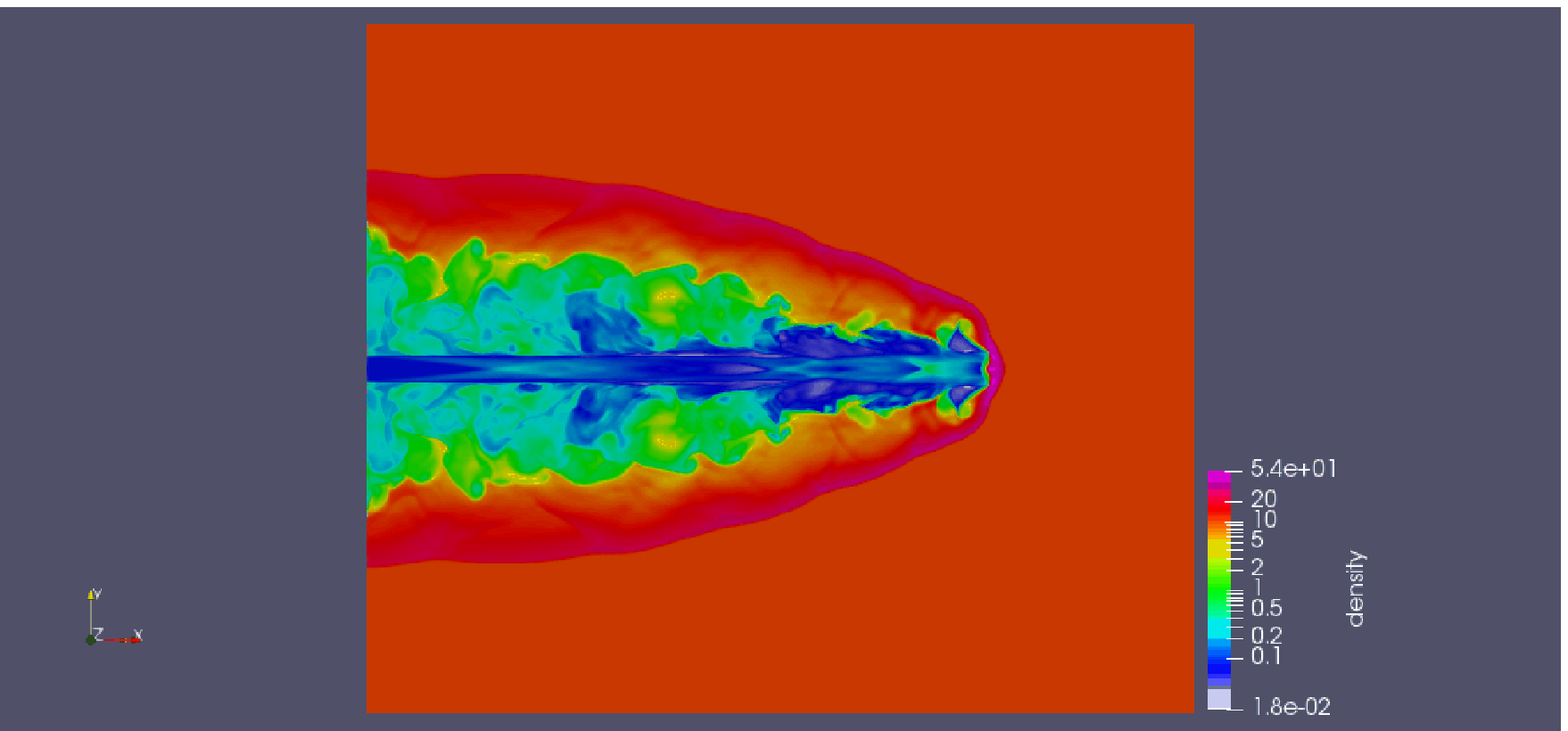}
\includegraphics[trim=4cm 0cm 2cm 4.4cm,width=0.45\textwidth,clip]{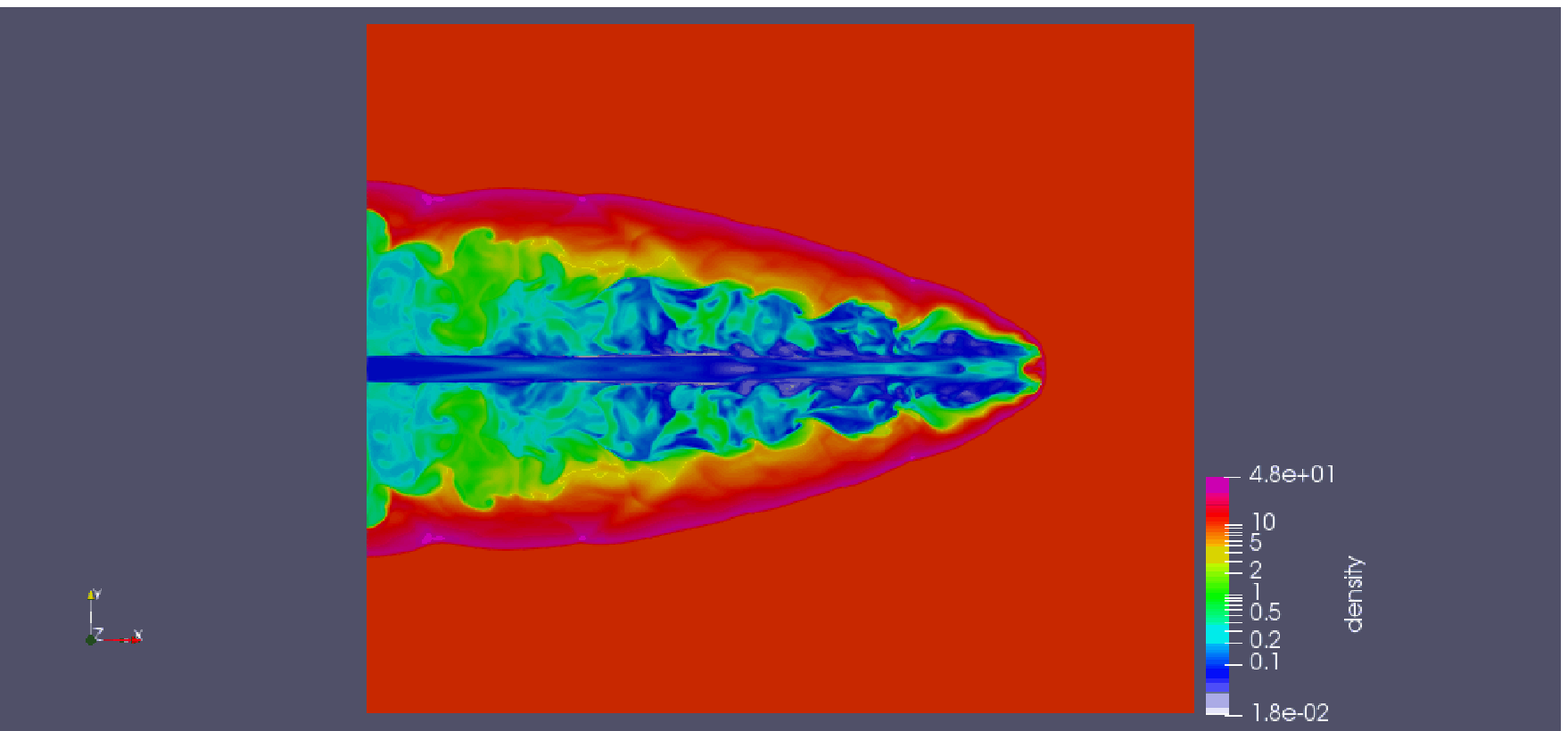}
\end{minipage}
\caption{Comparison between standard (left, $P_{j} / P_{a} = 1$) and overpressured (right, $P_{j} / P_{a} = 10$) simulations close to 60 $R_{j}$.
Top row: $v_{j} =0.96c$, $\eta$ =0.02 ($t=169$); middle: $v_{j} =0.98c, \eta$ =0.02 ($t=118$); bottom: $v_{j} =0.99c, \eta$ =0.0075 ($t=120$).}
\label{fig:overpressure}
\end{figure*}

It is quite clear that the jets in cases c1 and c2 terminate further and further behind the bow shock once they go fully unstable; at this time ($t\simeq400$ for c1 and $t\simeq350$ for c2) the terminal location of the jet,  which is the strongest shock within the flow, essentially stalls.  Nonetheless additional energy, momentum and mass continue to flow through the jet into the cocoon, which still has sufficient pressure  for some time to inflate a bow shock.   However, because those inputs are spread over a widening cocoon the rate of shock advance slows substantially  and may eventually decay into a weaker bow wave.  For runs c3--c9 the terminal jet shocks  which continue to be the strongest within the jets (and would be observed as hot spots) remain very close to the bow shocks throughout the simulations. The rate of advance of the bow shock remains remarkably constant for nearly all of these FR II like runs, although the least powerful of these (c4) actually accelerates at $t \simeq 240$, where the bow shock narrows and allows a more focused thrust.

In summary, unstable FR I type morphologies are produced by slower and/or lower density jets while faster/higher density ones yield FR IIs. The critical variable is the jet power and from  Table 1 we see that (at least out to the distances we simulated) and for the particular assumptions about constant jet radii and ICM properties we considered, the boundary between FR I and FR II sources is around $L_j \approx 1.2 \times 10^{46}$ erg s$^{-1}$.

Unfortunately, there are no iron-clad ways to use observations to constrain $L_j$ values from the measured radio fluxes (for FR IIs) or X-ray cavity measurements (particularly helpful for FR Is) but  attempts to make such connections yield lower values than those given in our Table 1.  For instance, \citet{Godfrey13} estimate jet power values for a substantial sample of FR II sources to range between $\sim 1 \times 10^{44}$ and $\sim 2.4 \times 10^{46}$ erg s$^{-1}$ while \citet{Cavagnolo10} estimate those for FR I sources to range between $\sim 4 \times 10^{40}$ and $\sim 8 \times 10^{45}$  erg s$^{-1}$.   A reasonable boundary between these types appears to lie around $10^{45}$ erg s$^{-1}$ \citep{Godfrey13}.  Our simulations have focused  on the higher power sources that might yield FR IIs, so  even lower $\eta$ and $v_j$ values than those we considered would certainly yield FR I types of much lower power, even if we kept the physical quantities to which we were scaling our simulations the same.   In addition, we could have certainly chosen lower, but still reasonable, values for $R_j$ or $\rho_{ICM}$ with which to scale our simulations and doing so would  have improved the match with these observational estimates.
We note that there is a large spread, up to a factor of $\sim 100$, in the jet powers that are estimated to produce the same observed radio flux for FR I sources, while the relation for FR II sources is tighter, but still spans nearly a decade \citep{Cavagnolo10,Godfrey13}.

\cite{Mass16} recently performed several  3D HD simulations of supersonic jets propagating through media with declining densities and pressures to investigate the morphologies of low power jets. They showed that the jet energy in the lower power sources, instead of being deposited at the terminal shock, was gradually dissipated through turbulence. The jets spread out while propagating, and they smoothly decelerate while mixing with the ambient medium and produce the plumes characteristic of FR I objects.  These simulations confirm the early analytical work  that showed that propagation on modestly relativistic jets through a declining atmosphere (usually on scales of $1-30$ kpc) should induce a jet slow-down to transsonic speeds and a substantial increase of the jet opening angle \citep{Gopal91,Bick94,Bick95}.  While these effects are almost certainly important and might help explain the discrepancy between the power boundary for our FR I and FR II simulations and the lower one estimated from observations, we have here shown that FR I type sources nonetheless can form for relatively weak relativistic jets, even in ambient media taken to have constant density and pressure.

\begin{figure*}[!htbp]
\begin{minipage}{\textwidth}
\centering
\includegraphics[trim=4cm 0cm 2cm 4.45cm,width=0.45\textwidth,clip]{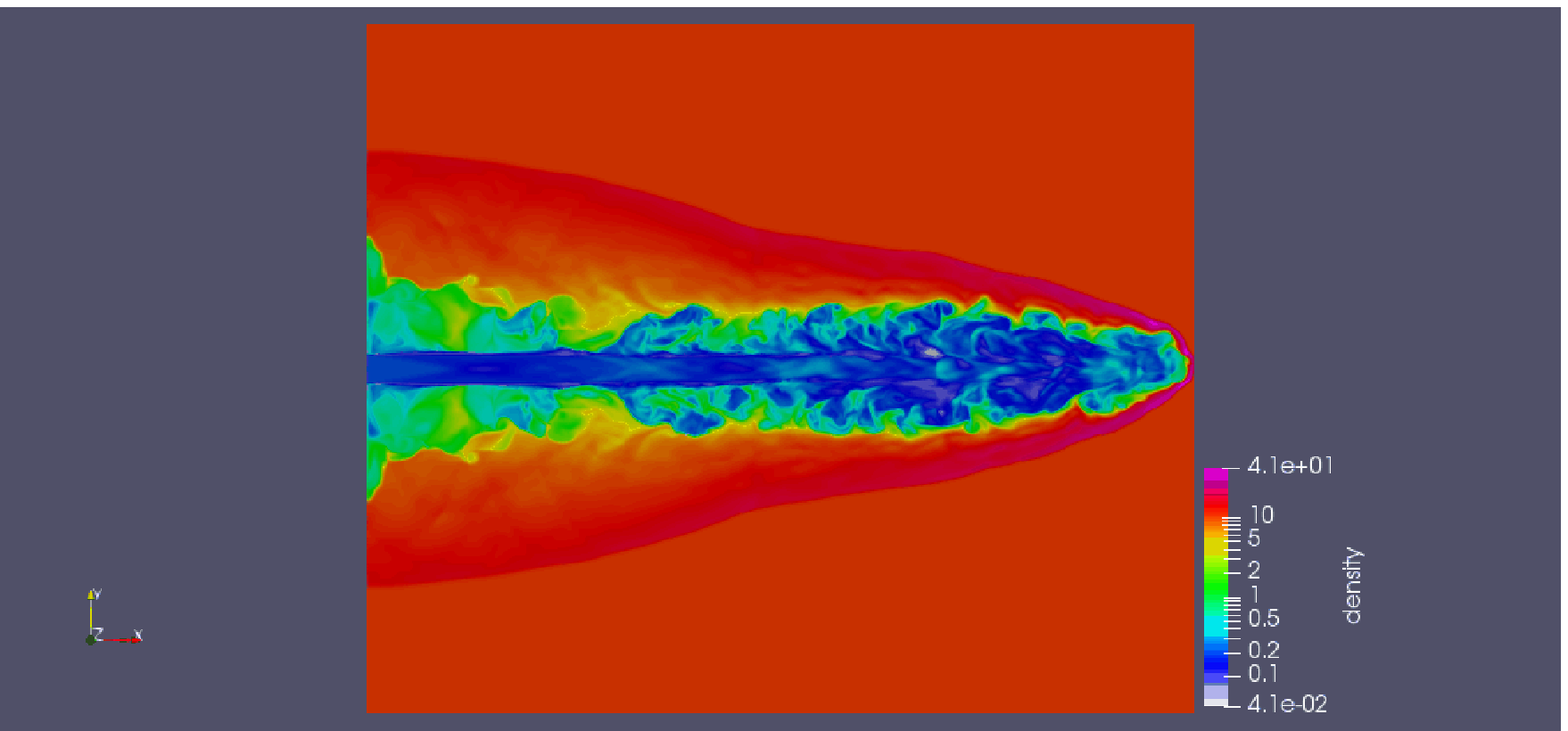}
\includegraphics[trim=4cm 0cm 2cm 4.45cm,width=0.45\textwidth,clip]{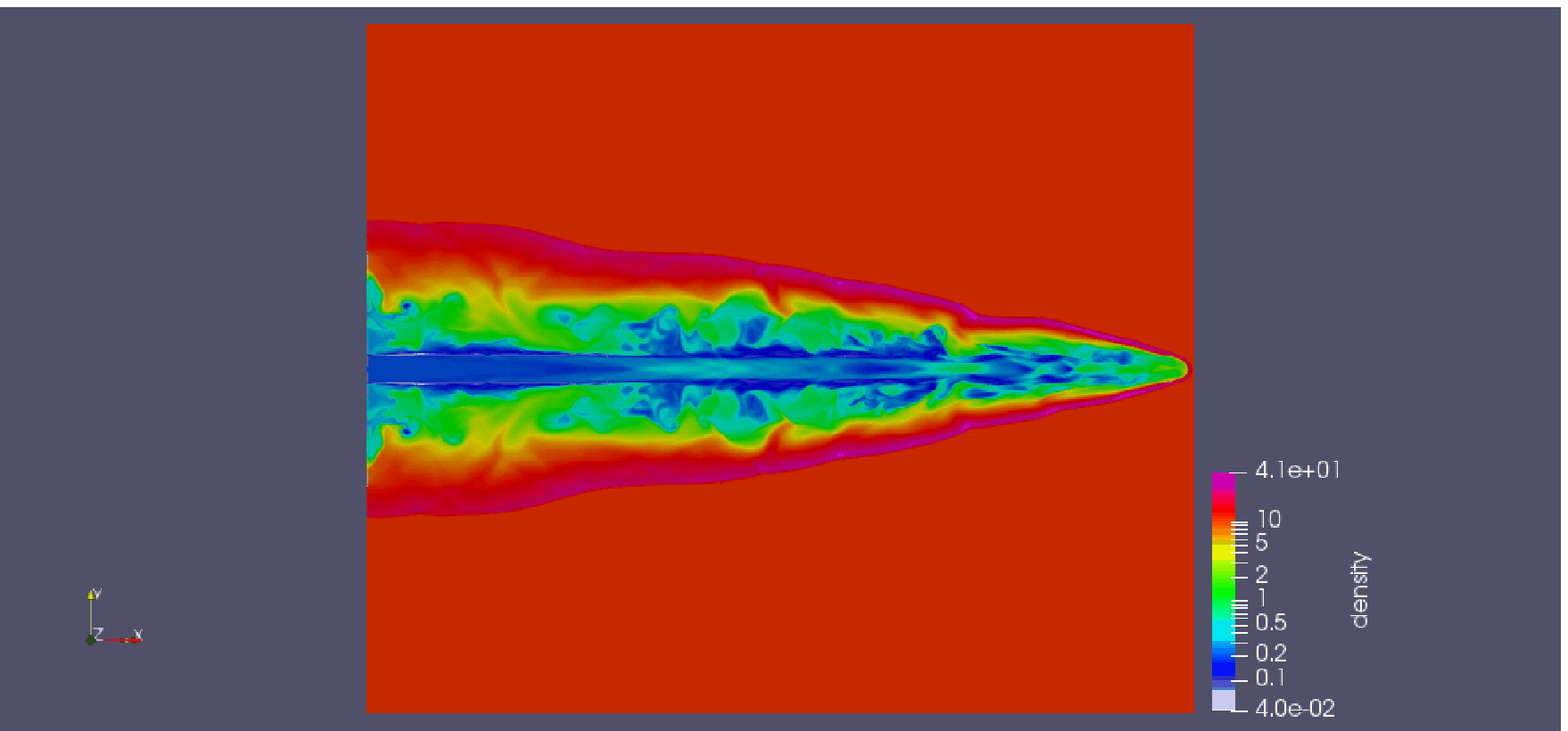}
\includegraphics[trim=4cm 0cm 2cm 4.45cm,width=0.45\textwidth,clip]{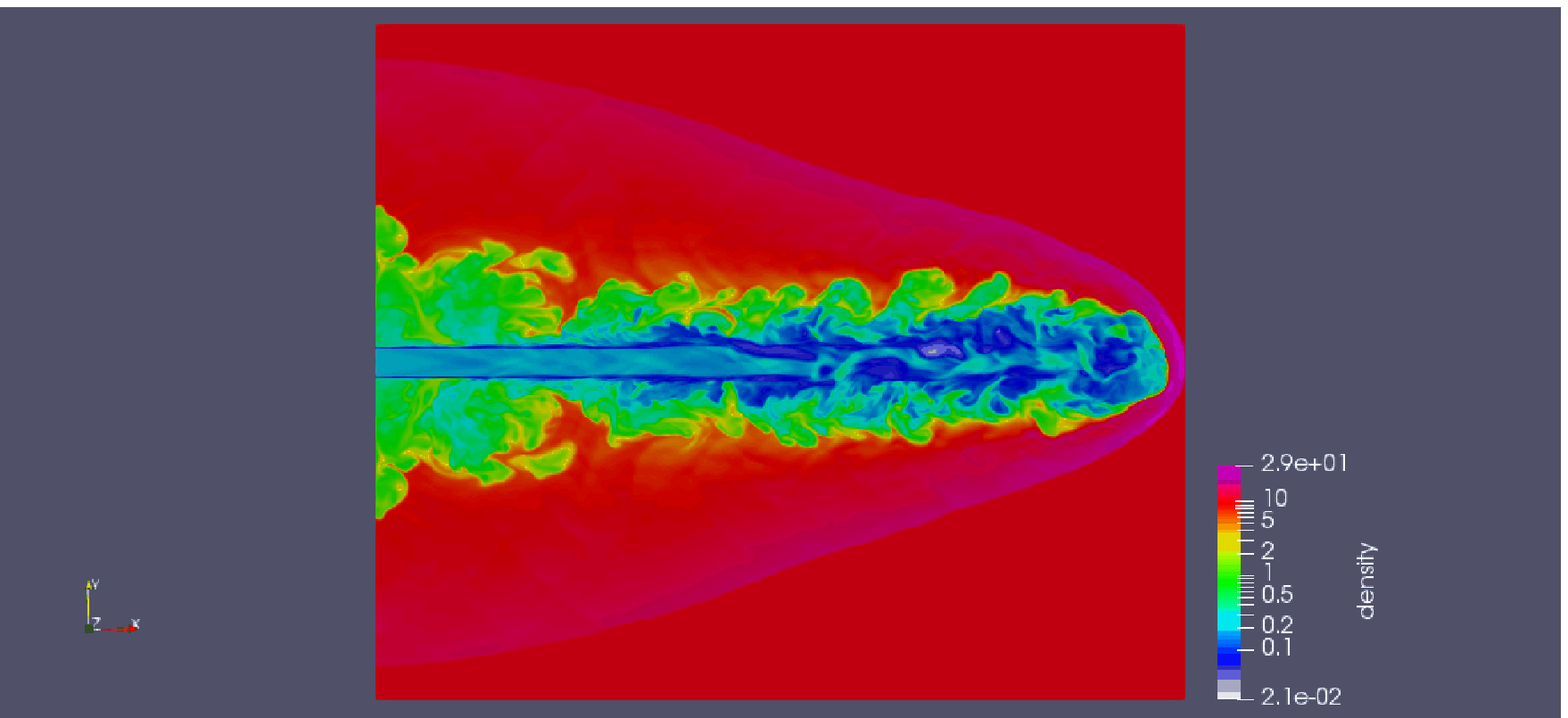}
\includegraphics[trim=4cm 0cm 2cm 4.45cm,width=0.45\textwidth,clip]{fig4c.eps}
\end{minipage}\vspace{0.1cm}
\caption{Density slices for $z = 0$ of simulations when the bow-shocks reach $60 R_j$ for  $\Gamma = 4/3$ in the upper row and  $\Gamma = 5/3$ in the lower row: (left) $v_{j} =0.85c$, $\eta$ =0.02 ($t=252, 335$);   (right) $v_{j} =0.96c$, $\eta$ =0.02 ($t=116, 170$). }
\label{fig:lowgamma}
\end{figure*}

\subsection{Overpressured simulations of powerful jets}
In all of our simulations mentioned so far (as well as those in the vast majority of the jet simulation literature cited in Section 1) it has been assumed that the jet is initially in pressure equilibrium with the ambient medium, with the ratio of jet pressure and ambient pressure $P_{j} / P_{a} = 1$. This assumption is motivated by the fact that an overpressured jet develops a strong recollimation shock that leads to the development of equal pressures between the internal and external gas \citep[e.g.][]{Belan10}. This equilibration probably occurs at radii within the initial 100 pc of the jet length and so assuming that the jet is already in pressure equilibrium at injection is quite sensible for sources propagating through the ICM or an even lower density IGM.  However, on smaller scales, the jets may well be overpressured and so we have considered three such runs which were simulated with $P_{j} / P_a = 10$, and have been marked with asterisks in Table \ref{tab:parameter}.

Density snapshots at late times for these overpressured simulations are shown in the right column of Figure \ref{fig:overpressure}, where  the equal pressure runs with the otherwise identical input parameters are displayed in the left column of Figure \ref{fig:overpressure} at the same times.  In all cases the morphologies of the simulation are similar, particularly for the bow shocks and cocoons. There are, however, a few modest differences that can be observed: the overpressured simulations  propagate a slightly longer distance  than do the equal pressure ones at the same times, their jets are wider close to the inlet and their first recollimation shock occurs further down the jet.  All of these differences would be expected because of the higher jet pressures.

\subsection{Effects of varying adiabatic index}

An optimal treatment of the equation of state (EoS) in RHD simulations would allow for smooth transitions between fully relativistic jet fluids, where the adiabatic  EoS, $\Gamma = 4/3$, is a good approximation, and the hot, but still essentially classical, fluid in the unshocked ambient medium where $\Gamma = 5/3$ is highly accurate.  Various computationally efficient approximations to the complex analytical \citep{Synge57} fully relativistic EoS for non-magnetized  fluids have been proposed \citep[e.g.][]{Mignone05,Ryu06} and this transitional EoS approach was extended to RMHD flows by \citet{Mignone07}.  Unfortunately, the Athena code is not designed to accommodate different EoSs for the different fluids, nor can it currently handle a transitional EoS.  Hence we have computed three simulations where $\Gamma = 4/3$ is assumed throughout to see if consistent differences arise in the results for a different adiabatic EoS.

The full Synge EoS was used in \citet{Scheck02} where they analyzed three 2D RHD jet simulations with different compositions (leptonic or baryonic) and found very little difference between the morphologies and dynamic behaviors of those light ($\eta = 10^{-3}$ or $\eta = 10^{-5}$) jets.
\citet{Mignone07} presented a different EoS comparison, also for 2D RHD simulations, for one very fast ($\gamma_j = 10$) and low density ($\eta = 0.001$) set of simulations where they found the $\Gamma = 4/3$ jet to propagate substantially faster than the $\Gamma = 5/3$ one.  They also computed a transitional EoS simulation which, unsurprisingly, led to results in between those  of the two fixed $\Gamma$ values.

 In Figure \ref{fig:lowgamma} we display late-time density slices of two of our 3D RHD high-resolution runs where we  used $\Gamma = 4/3$ in the upper row and $\Gamma = 5/3$ in the lower row.   The first run ($v_j = 0.85c, \eta = 0.02$) has become unstable by the time the jet goes somewhat beyond $30 R_j$, while the second run ($v_j = 0.96c, \eta = 0.02$) remains stable out to at least $60 R_j$.   There are no huge differences between the $\Gamma = 4/3$ and $\Gamma = 5/3$ simulations in terms of basic shape and stability, as was also seen for a simulation with $v_j = 0.98c, \eta = 0.01$ which we do not display.  However, there are interesting distinctions.  These 3D simulations with $\Gamma = 4/3$ move faster across the grid by  modest factors of $1.3-1.5$ than those with $\Gamma = 5/3$.  This greater advance speed implies stronger and narrower bow-shocks and slightly narrower cocoons when compared at equal distances (as shown in Figure \ref{fig:lowgamma}), but when compared at equal times, while the bow-shocks remain narrower, the cocoons, at least in the outer-halves of the sources, are of similar width.    It is worth noting that the differences between jets with alternative adiabatic indices in our 3D simulations are in the same senses as those seen earlier for 2D jets \citep{Scheck02,Mignone07}, but are not as pronounced in 3D.  It also appears that the jets in the weaker $\Gamma = 4/3$ simulations may remain stable for modestly longer distances and times.

\subsection{Comparison of RHD jets in 3D and 2D}
\begin{figure*}[!htbp]
\begin{minipage}{\textwidth}
\centering
\includegraphics[trim=3cm 0cm 2cm 4.4cm,width=0.48\textwidth,clip]{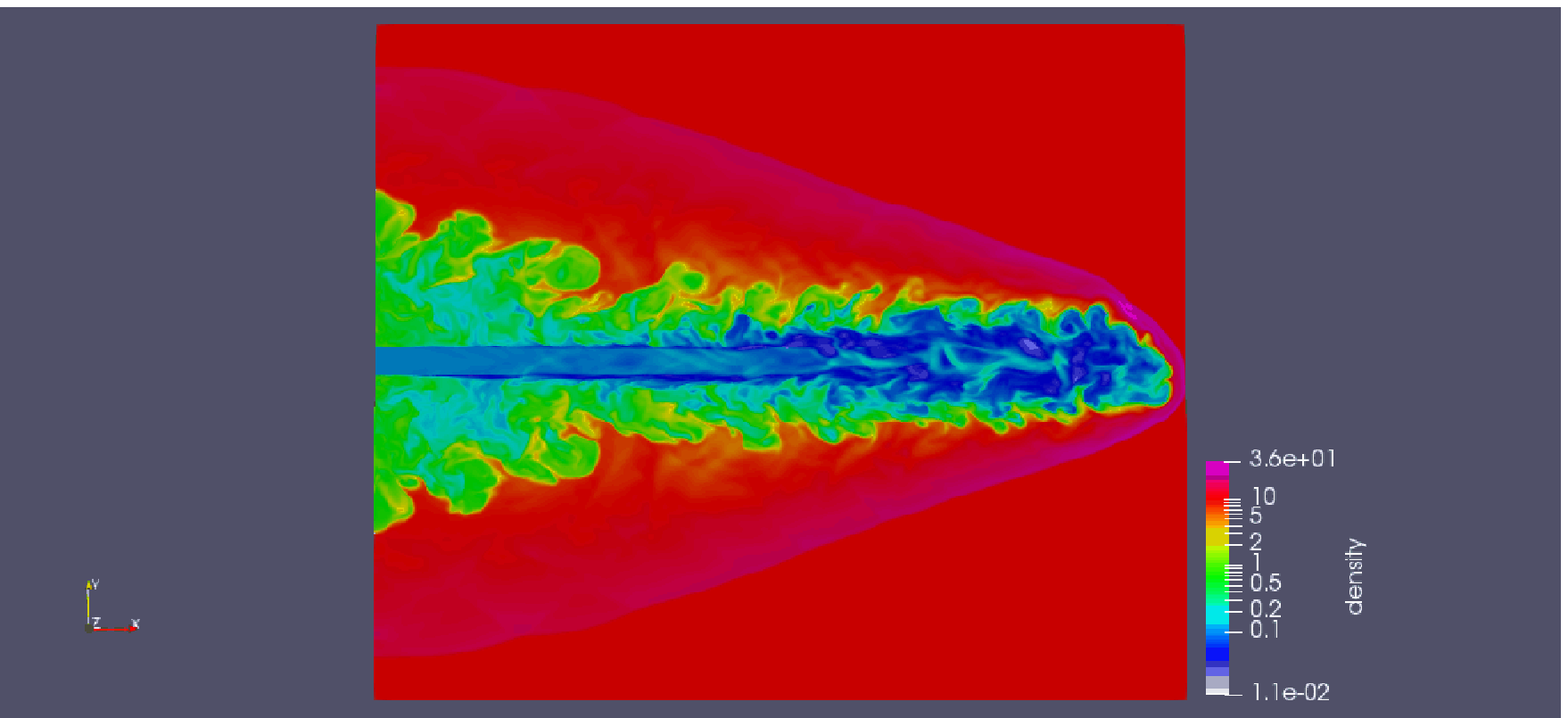}
\includegraphics[trim=3cm 0cm 2cm 4.4cm,width=0.48\textwidth,clip]{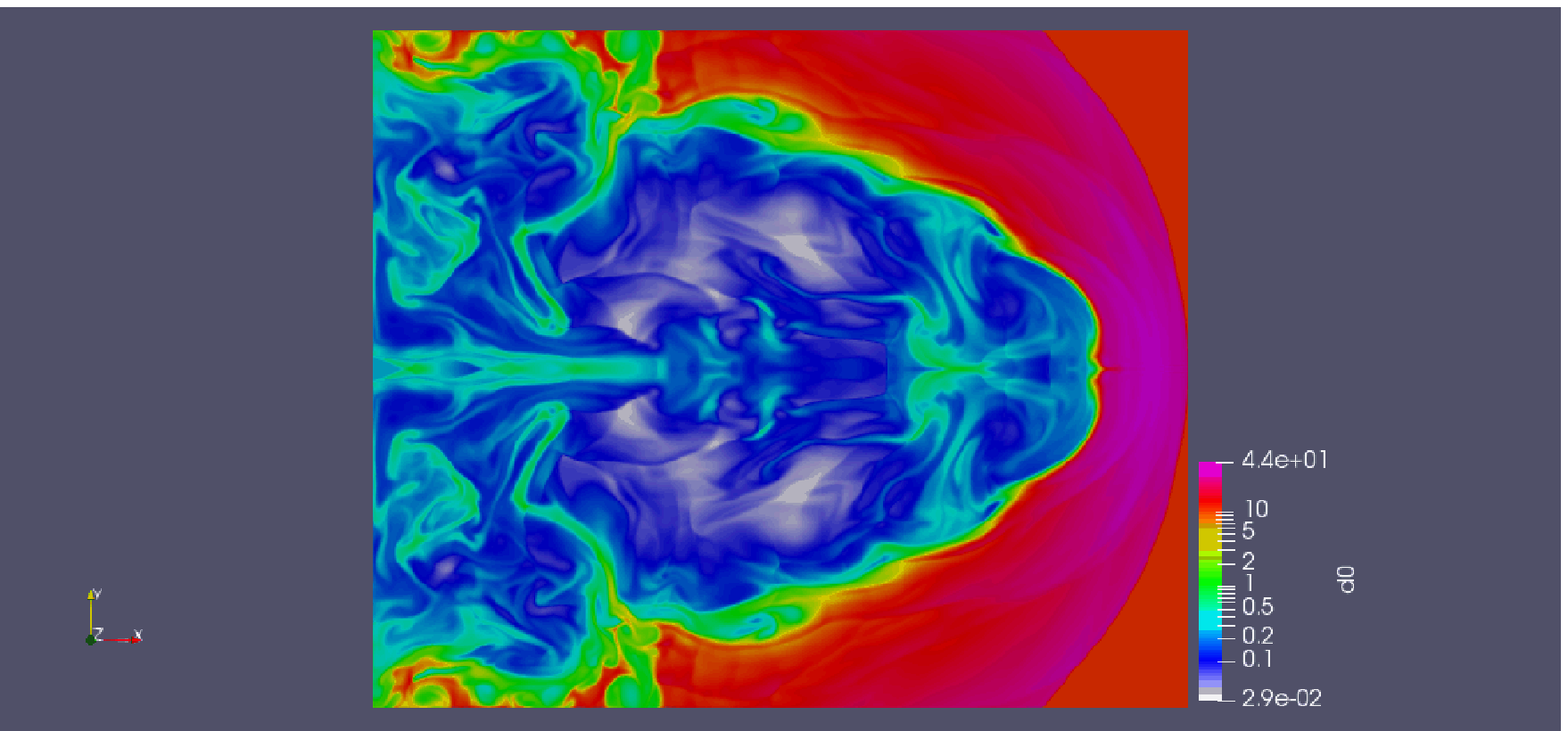}
\end{minipage}\vspace{0.1cm}
\begin{minipage}{\textwidth}
\centering
\includegraphics[trim=3cm 0cm 2cm 4.4cm,width=0.48\textwidth,clip]{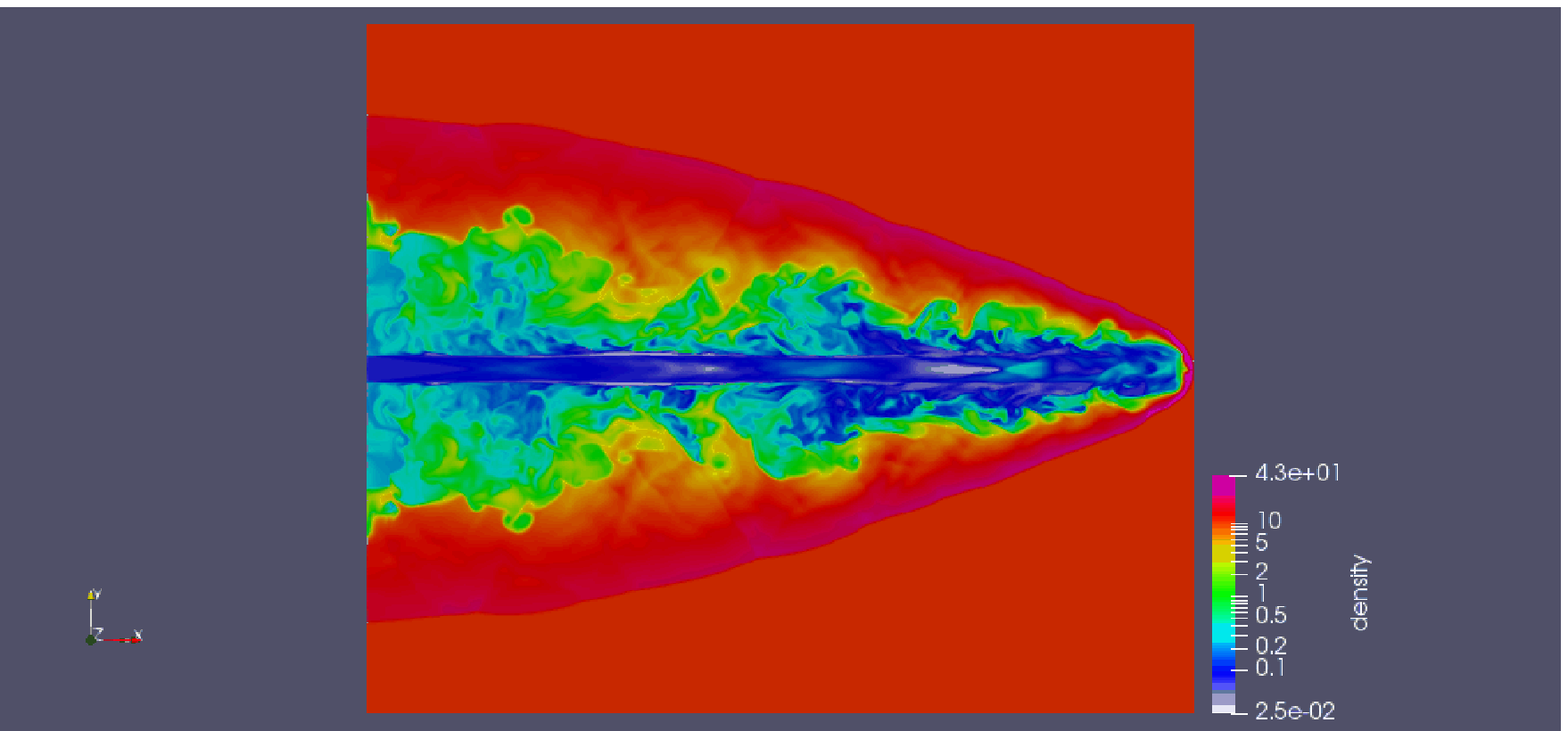}
\includegraphics[trim=3cm 0cm 2cm 4.4cm,width=0.48\textwidth,clip]{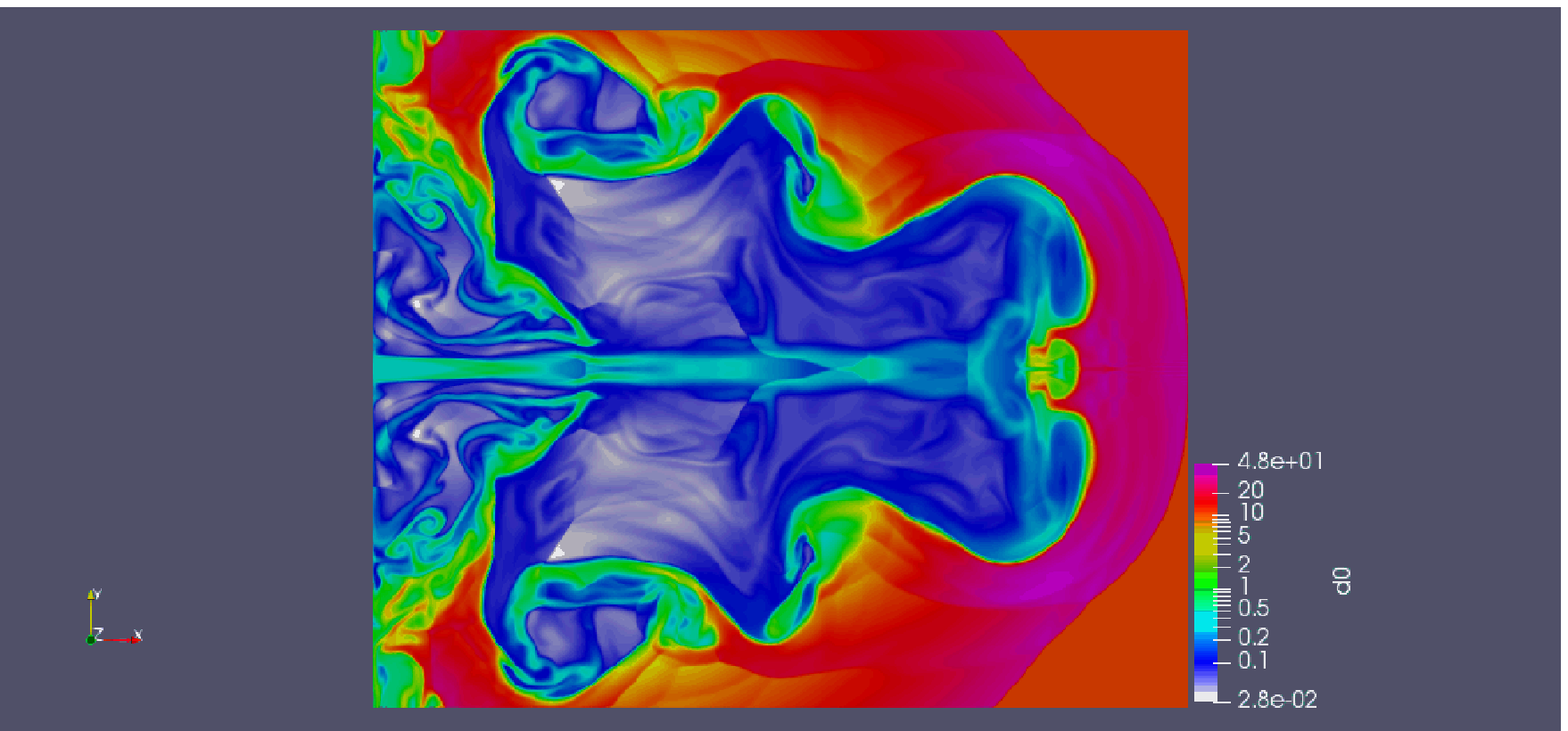}
\end{minipage}
\caption{High resolution RHD runs for 3D and 2D slabs at the same distances:
(top-left) 3D simulation with $v_{j}=0.93c$, $\eta$ = 0.01 at $t=294$; (top-right) 2D with the same parameters for $t=367$; (bottom-left) 3D with $v_{j}=0.98c$, $\eta$ = 0.0075, $t=189$; (bottom-right) 2D with the same parameters, $t=233$.}
\label{fig:2Dplot}
\end{figure*}

Our work here on 3D RHD simulations follows our previous 2D RHD relativistic jet simulations \citep{Poll16} which were focused on the variability induced by jet motions. Even though the 3D runs are clearly superior and were enabled by The College of New Jersey's (TCNJ's) recent acquisition of a large computer cluster, it should be worthwhile to compare the morphology of jets produced in 2D and 3D RHD simulations using the Athena code at the same high resolutions.  As shown in Figure \ref{fig:2Dplot}, there are major differences between 3D (left) and 2D-slab (right) simulations regardless of input parameters. The top row shows a weaker jet that goes unstable rather quickly and the lower one a  more powerful jet that remains stable at the times when their bow shocks reach the end of the grid at $60 R_j$.   This critical distinction is seen in both 3D and 2D simulations, with the jet going unstable at roughly the same distance in both of the lower-power cases. The morphology of 2D simulations is more symmetric than 3D ones as fewer instabilities can be excited in the former. The big difference is that 2D simulations inflate much wider bow shocks and cocoons and thence lose material off the grid along the upper and lower boundaries.  The jet widths are similar, though slightly wider for the 2D runs at large distances.  Concomitantly, 2D simulations  take longer times to cross the entire grid and their jet termini are much further behind  the bow shock than they are in the 3D simulations.

\section{Light curves and power density spectra}

\begin{figure}
   \begin{minipage}{\textwidth}
   \centering
   \includegraphics[trim=0cm 0cm 1cm 0cm,width=0.33\textwidth,clip]{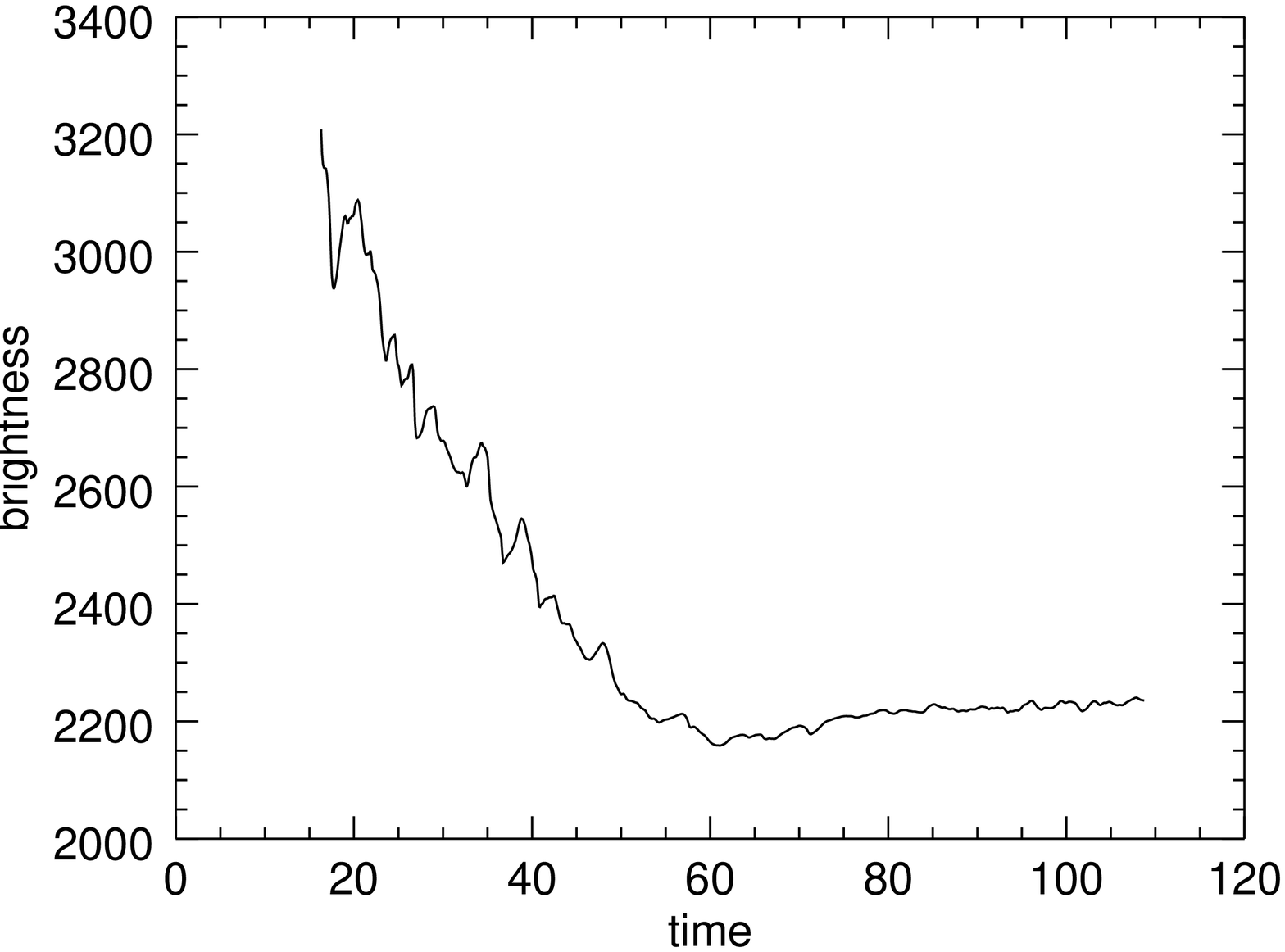}
   \includegraphics[trim=0cm 0cm 1cm 0cm,width=0.33\textwidth,clip]{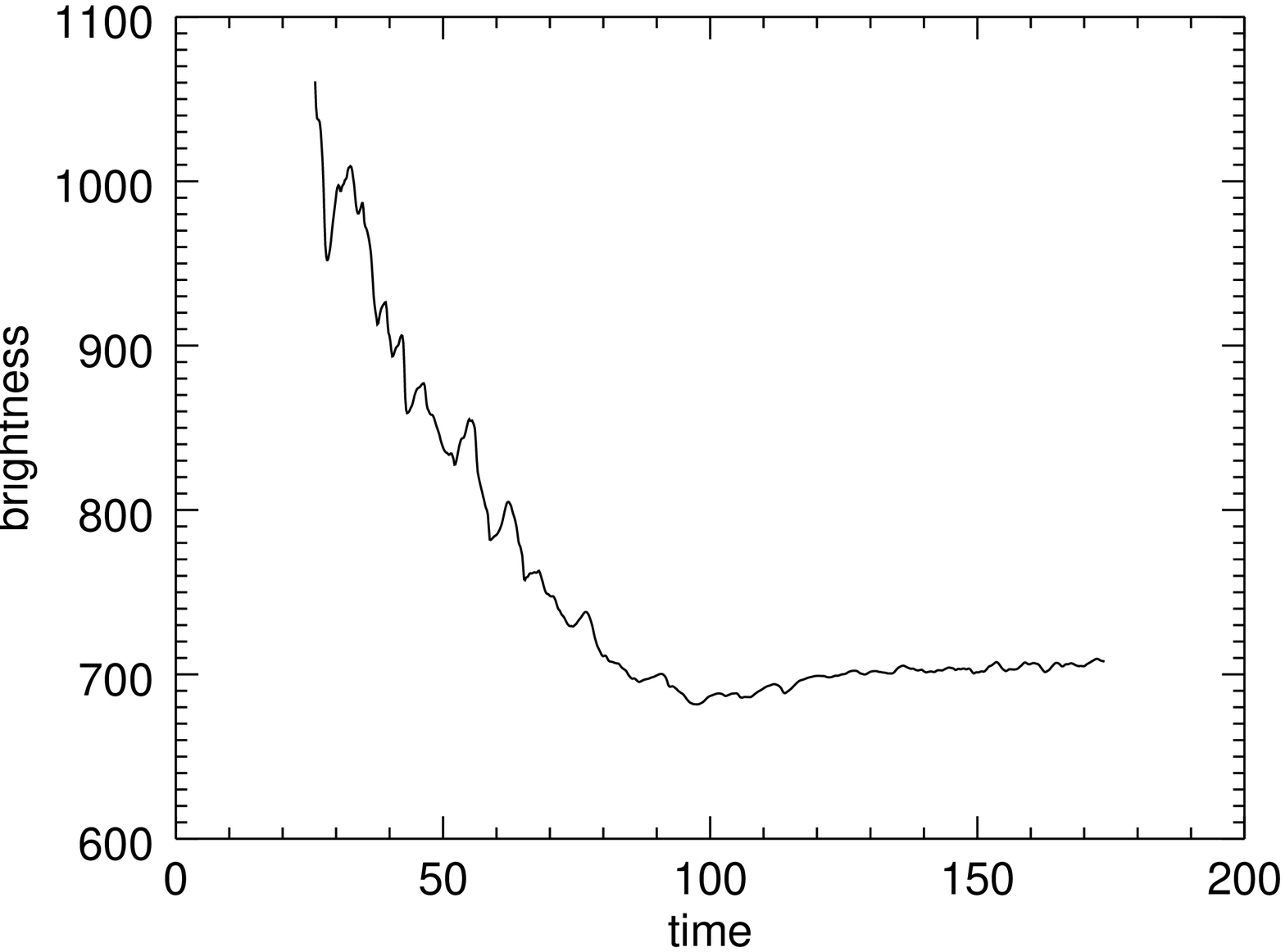}
   \includegraphics[trim=0cm 0cm 1cm 0cm,width=0.33\textwidth,clip]{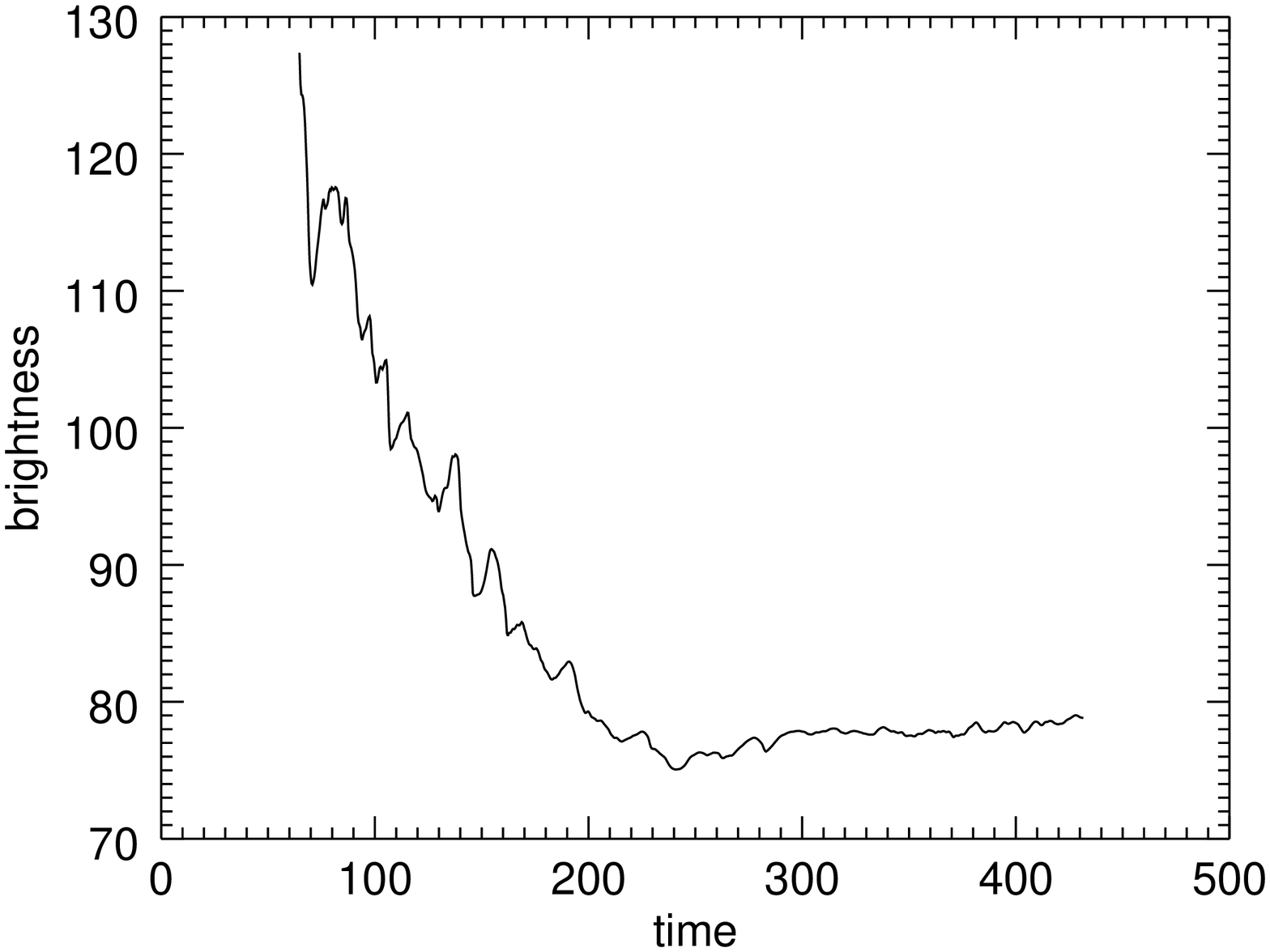}
   \end{minipage}\vspace{0.001cm}
   \caption{Light curves for $v_{j} = 0.985c$, $\eta$ = 0.02 at different viewing angles:
   (left)  $\theta = 5^{\circ}$; (middle) $\theta = 10^{\circ}$;  (right) $\theta$ = $20^{\circ}$.}
   \label{fig:3viewangle}
\end{figure}

A key new aspect of this work is our consideration of the variations of emission produced by 3D propagating relativistic jets.  To encompass variability time-scales of practical interest, i.e., those that are shorter than decades, we must scale the jets downward so that they correspond to jets emerging from the vicinity of the central engine where the length scales are parsec-like and not kpc-like and where Doppler boosting shortens the observed times of variations when compared to those occurring in the rest frame of the jets. Our viewing angles to the relativistic jets in AGN we call blazars are  quite small and so we have usually taken $\theta = 10^\circ$ in computing crude light curves as described in Section 4.  Observed light curves for an extended run (to $t =1000$ in the source rest frame) with $v_{j} = 0.985c, \eta = 0.02$ are shown in Figure \ref{fig:3viewangle} for different viewing angles:  $\theta = 5^\circ, 10^\circ$ and $20^\circ$.  Smaller viewing angles produce higher observed fluxes over the shorter periods in the observer's frame. The emission for this case when viewed at $\theta = 5^\circ$ has an amplitude about 25 times higher than that seen at $\theta = 20^\circ$ thanks to the higher boosting factor.

\begin{figure*}
    \centering
    \includegraphics [trim=0.2cm 0cm 0.2cm 0cm,width=0.8\textwidth,clip]{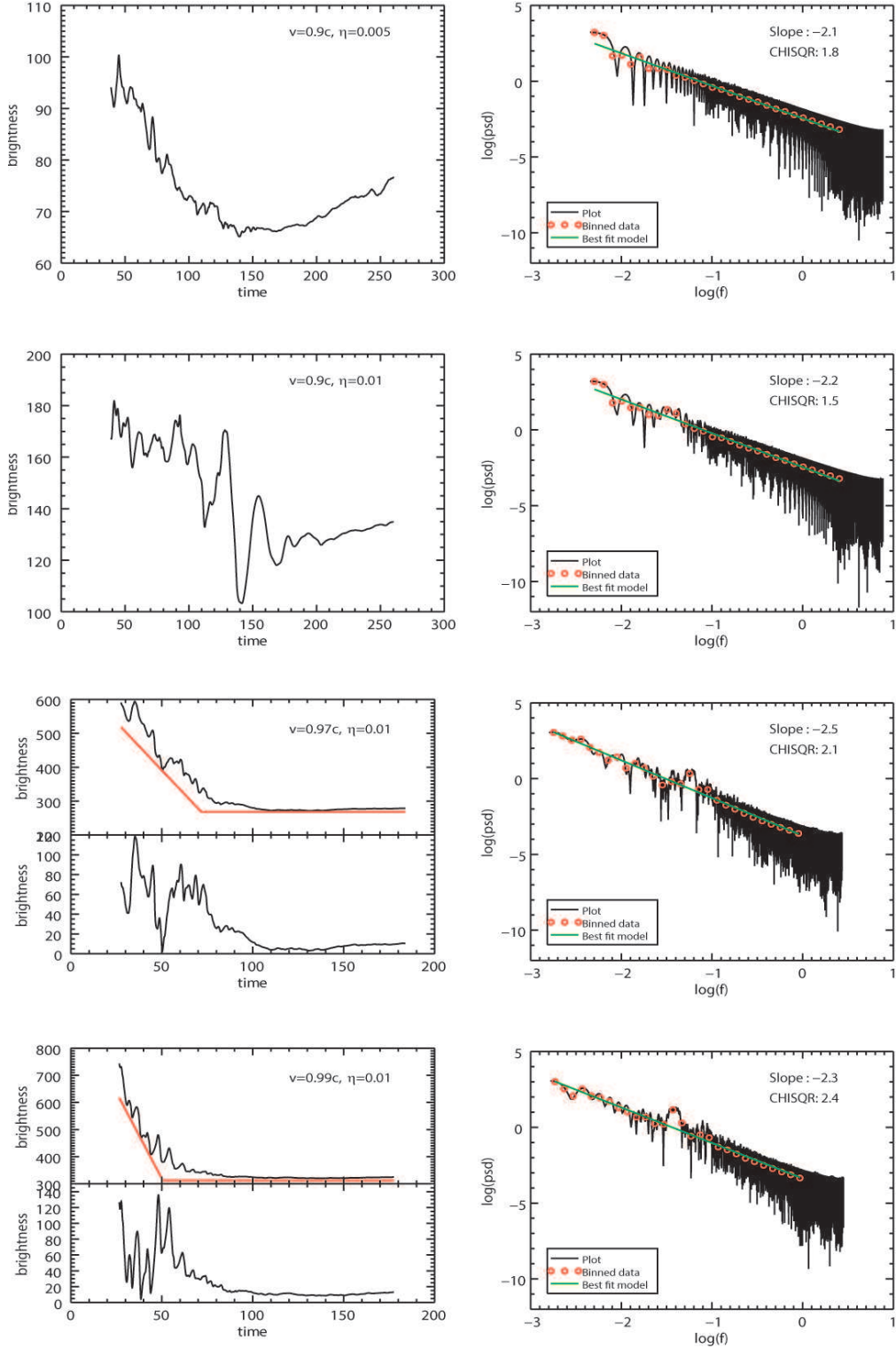}
    \caption{Observed light curves (left) and PSDs (right) at viewing angles of $10^\circ$, where parameters and slopes are marked on the panels. The light curves for the bottom two simulations ($v_{j} = 0.97c $ and $v_{j} = 0.99c$) are shown in two ways: upper sub-panels with full observed emission variability in black and the long-term trend baseline components underlying the variability in red;  the bottom sub-panels show the light curves after subtraction of these best-fit trend lines. In the right column images the (red) circles are binned data points and (green) lines are the best fits yielding the quoted low-frequency slopes; PSDs computed from the post-subtracted trends are shown in the bottom two panels.}
   \label{fig:psd}
\end{figure*}

Using the nuclear scaling ($R_j = 1$ pc) the rest frame time unit is   3.26 yr and the interval at which the light curves were sampled is $\Delta t = 0.815$ yr.  For the range of bulk Doppler factors considered in Figures \ref{fig:3viewangle} and \ref{fig:psd},  of 2.3 to 9.2, this means that time intervals down to 0.089 yr (essentially one month) are being simulated.  The nominal intrinsic powers for such jets moving through the galactic nuclear region lower by a factor of $4 \times 10^{-3}$ than those quoted in Table 1, where they were scaled for propagation through the ICM, yielding a rescaled range from $2.3 \times 10^{42}$ erg s$^{-1}$  to $2.3 \times 10^{45}$ erg s$^{-1}$.  However, here we are fundamentally  concerned with seeing if the fractional levels of variability produced by changes in the Doppler boosting of the different cells of the jet in the region close to the first reconfinement shock, and the resulting power-spectra, are in reasonable agreement with those seen in blazars. We note that the values of $R_j$ and $\rho_{nuc}$, which set the nominal $L_j$ through Eqn.\ (5), can all have significant ranges about our nominal values.

 For several emission light curves, all explicitly evaluated at a viewing angle of $\theta = 10^\circ$, the power spectral density (PSD) were computed using a Lomb-Scargle periodogram, which is a method to estimate the PSD generated by a physical process that can sensibly deal with unevenly sampled time series data \citep{Horne86}, although they are computed at equal intervals here.  The periodogram is evaluated using the algorithm presented in \cite{Press89} in order to achieve fast computational speeds.
 The observed PSDs of AGNs have either been measured in X-ray bands for Seyfert galaxies or in optical bands for quasars and blazars.  In either case, they are usually found to be approximated by red noise at lower frequencies, with the power depending on frequency roughly as $P \propto f^{\alpha}$ with $\alpha \approx -2$, while at high frequencies, measurement white noise ($\alpha \simeq 0$) dominates; in a small minority of cases, a flatter slope, $\alpha_{low} \approx  -1$, may be present below a relatively low  break  frequency \citep{Uttley05,Gonz12,Edel13,Wehrle13,Revalski14,Smith18}.

Four examples of our simulated emission light curves (left) and their corresponding PSDs (right) are shown in Figure \ref{fig:psd}.  For the first two cases, both with the same value of jet velocity ($v_{j} = 0.9c$) but different jet densities, these light curves show overall substantial variations (55--75\%) with some rapid fluctuations.   During the first 150 (proper) time units of the 1000 for which these were run, the bow-shock and terminal shock move past the portion of the grid that we analyze (which is shown in Fig.\ 6), so we do not consider that portion of the light curve. The decreasing fluxes seen in most of the early portions of the simulations as plotted arise from the transient effect of the establishment of the the reconfinement shock, whose position and strength often takes a while to stabilize. After that phase, the variations are more modest (on the several percent level) and reflect the changes in the velocities, densities and pressures of the relevant jet zones once a quasi-steady state is established in the inner portion of the jet.  But there are stronger longer-term variability trends for the faster jets shown in the lower two panels ($v_{j} = 0.97c$, and $v_{j} = 0.99c$), both for the same $\eta$.  In both of these cases the baselines can be considered to be  composed of two parts: a declining trend for the first 20--30\%, and a constant one for the remaining portions of the light curves.   Before proceeding to compute the PSDs of the emissions, it probably makes sense to subtract those baseline trends, but we have computed the PSDs with and without subtracting the trends and give the results both ways in Table 3.  The top panels of the $v_{j} = 0.97c$ and $v_{j} = 0.99c$ simulated light curves show the full light curves and the baselines while  the bottom panels show the light curves after subtraction of the best-fit line trends for each simulation.

 \begin{table*}
    \centering
    \begin{threeparttable}
    \centering
    \caption{PSD results}
    \label{tab:psd}
    \begin{tabular}{clccc}
    \hline\hline
      $\beta_{j}$ & $\gamma_{j}$ & $\eta$ & $\alpha$ & $\chi^2_r$ \\\hline
       0.9          & 2.29         &0.005   & $-$2.1  & 1.8    \\
       0.9          & 2.29         &0.01    & $-$2.2  & 1.5    \\
       0.9          & 2.29         &0.02    & $-$2.2  & 1.8    \\
       0.9\tnote{*} & 2.29         &0.02    & $-$2.4  & 1.5    \\
       0.95         & 3.20         &0.01    & $-$2.2  & 1.4     \\
       0.95\tnote{*}& 3.20         &0.01    & $-$2.4  & 3.3    \\
       0.95         & 3.20         &0.02    & $-$2.2  & 1.7     \\
       0.97         & 4.11         &0.01    & $-$2.2  & 1.4    \\
       0.97\tnote{*}& 4.11         &0.01    & $-$2.5  & 2.2    \\
       0.97         & 4.11         &0.02    & $-$2.2  & 1.9    \\
       0.97\tnote{*}& 4.11         &0.02    & $-$2.4  & 2.8    \\
       0.985        & 5.80         &0.02    & $-$2.2  & 1.6    \\
       0.985\tnote{*}& 5.80        &0.02    & $-$2.3  & 1.5    \\
       0.99         & 7.09         &0.01    & $-$2.1  & 1.9    \\
       0.99\tnote{*}& 7.09         &0.01    & $-$2.3  & 2.4    \\
    \hline
        \end{tabular}
\begin{tablenotes}
\item[*] A linear trend was subtracted from the first portion of the emission light curve before the PSD was computed.
\end{tablenotes}
\end{threeparttable}
\end{table*}

The right column of Figure \ref{fig:psd} shows the corresponding PSDs of these light curves with those for which the trends were subtracted being displayed for the bottom two panels.   To compute the PSDs, we adopted Hanning window functions which reduce red-noise leakage in Fourier transform filtering \citep{Har78,Mart14}. There are many fewer data points on the low frequency end of the  PSDs than on the high end. To mitigate this inequality, when computing the slopes of the PSDs we binned the data points at 0.1 dex frequency intervals and computed an average for each bin, which have been marked with red circles on the images. These averaged points for each bin were used to fit a power-law line to the data from the log--log plot of power against frequency to calculate a final slope up until the white noise contribution becomes important. Although each light curve is unique and has significant differences, we found a rather narrow range in the best fitting slopes of the PSDs.  The slopes and reduced $\chi_r^2$ values for the straight-line fits to the  red-noise portions of the PSDs that we have computed are listed in Table \ref{tab:psd} for 9 distinct runs, of which 5 were also calculated after trend subtractions were performed. The best fitting slopes of the resulting PSDs  had the range $-2.1 \ge \alpha \ge  -2.5$.  We note that the subtraction of a linear baseline from parts of the light curve produces a PSD slope that is steeper by 0.1--0.3 compared to the unsubtracted PSDs.  Regardless of approach taken, our PSDs are in accord with observations of optical and X-ray variations from AGN, which have been seen to range between $-1.5$ and $-3.4$ but do seem to cluster between $-2.0$ and $-2.5$ \citep{Mac10,Gonz12,Edel13,Revalski14,Kasliwal15,Mohan16,Smith18,Wehrle13,Wehrle18}.  Similar results ($-2.9 \le \alpha \le -2.1$) were obtained from our earlier and cruder 2D RHD simulations that did include the modest effects of taking the time-travel lags into account \citep{Poll16}.

It is certainly gratifying that these simulations produce light curves that are reminiscent of those of many blazars, in terms of having both active and quiescent periods and that the PSD slopes are also in accord with observations; however, we make no claim that this approach provides a unique explanation for those phenomena.  In particular, we note that very fast variations cannot plausibly be obtained by the changes in the properties of jet zones as we have  modeled them.  Activity on scales much smaller than the grid cells in our jet simulations appear to be necessary as well.  These would involve turbulence \citep{Mars14,Calafut15,Poll16} or ultrarelativistic mini-jets within the jets, most likely produced by magnetic reconnection \citep[e.g.][]{Giannios09,Biteau12,Kagan16,Orio17}.

\section{Conclusions}

We have simulated  an exceptionally large suite  of over 50 3D RHD propagating jets using the Athena code, with an emphasis on investigating their suitability for modeling observed blazar variability. Our simulations of propagating jets have spanned a significant range of velocities for the initial bulk flows ($0.7c - 0.995c$) that cover the great majority of the velocities deduced for radio galaxies and blazars \citep[e.g.][]{Lister09}. These flows are light, as is appropriate to radio jets, with jet density to ambient medium density ratios between $5.0 \times 10^{-4}$ and $3.2 \times 10^{-2}$. Both medium resolution (5 zones per jet radius) and high resolution (10 zones per radius) simulations have been completed extending out to 60, 120, or even 240 jet radii along the direction of motion; in all cases our simulations had widths of 50 jet radii in the two perpendicular directions so there was no loss of matter out of the grid along those transverse boundaries or the need to worry about waves reflecting off those boundaries and unphysically distorting the jet flow.

These simulations span a sufficient range in power so that the weaker ones go unstable before they pass through our simulation volumes.  When scaled to extragalactic dimensions and parameters expected for intracluster (or intergalactic) ambient media these cases yield FR I type morphologies; if scaled to dimensions and properties probed by VLBI, such dying jets would represent failed radio sources.  The majority of our simulations took more advantage of the relativistic velocities computable with Athena and correspond to powerful sources that remain stable for very extended distances and times.  On large scales these would be FR II radio galaxies and on the small scales they would be young radio galaxies, probably seen as compact steep spectrum sources if not aimed close to our line of sight.  Although bow shocks continue to be driven by the weaker sources for some time after the jets have clearly gone strongly unstable, the rate of propagation slows down, while our powerful relativistic jets remain extremely well collimated for very extended times and the rate of advance of both the bow shocks and the jet tips are close to constant for them.

The main suite of simulations was performed for  jets injected in pressure balance with the ambient medium and where an adiabatic index of 5/3 was taken for both jet and ambient fluids.  However, a few
 overpressured simulations ($P_{j}/P_{a} =10$) have been performed to compare with equal pressure simulations.  Their morphologies are quite similar, though unsurprisingly we saw that otherwise identical overpressured simulations propagate a little faster than do the standard ones and they have wider jets, at least at the early stages of the simulations.  We also performed a few simulations where $\Gamma = 4/3$ instead of 5/3 was assumed: the former propagated somewhat faster than the latter and may stay stable for slightly longer times but do not display major differences.  Comparison between our standard 3D RHD and 2D (slab-like) RHD simulations were also made and they show that 2D simulations have wider bow shocks and cocoons and take  longer to  propagate across the entire grid than do 3D ones.

 We approximate the synchrotron light curves of these propagating jets by summing up the fluxes emitted from zones with different densities and pressures, where the intrinsic emissivity is taken to be proportional to the product of $\rho$ and $P$.  As we deal with propagating jets, these physical quantities vary with time but are higher for more powerful jets.  Each of those rest frame ``emissivities'' is then differentially Doppler boosted by the changing velocities each zone evinces at each time-step downloaded from the simulations.  We then sum up these contributions over a substantial portion of the jet  close to the first reconfinement shock and then compute observed total brightnesses and their power spectra.  These do seem to be able to reproduce key aspects of blazar light curves and yield PSD slopes a little steeper than $-2$, which also are found to characterize a significant majority of observations of the optical and X-ray variations from AGN.

The biggest shortcoming of these simulations is that they have been restricted to flows where the magnetic fields have not been incorporated.  We are in the process of computing MHD simulations which we will then analyze in a similar fashion to see if those extra degrees of freedom (and complication) yield interestingly different variability characteristics.

\section*{Acknowledgements}
We thank the anonymous referee for useful criticisms and suggestions that have led to a substantially improved manuscript. The authors acknowledge use of the ELSA high performance computing cluster at The College of New Jersey for conducting the research reported in this paper and are most grateful to Shawn Sivy for assistance in porting the Athena code to the cluster and for his efforts in constructing and maintaining this facility. This cluster is funded by the National Science Foundation under grant number OAC-1828163.
ParaView software was used for the production of most of the figures. YL was supported in part by the China Scholarship Council (file No.\ 201706220272).  This work is partly supported by Natural Science Foundation of China under grant No.\ 11873035, the Natural Science Foundation of Shandong province (No.\ JQ201702) and the Young Scholars Program of Shandong University (No.\ 20820162003).

\end{document}